\documentclass[reqno,11pt,article]{amsart}
\usepackage{latexsym}
\usepackage[a4paper,left=25mm,right=25mm,top=25mm,bottom=25mm,marginpar=25mm]{geometry}

\usepackage{graphicx}

\usepackage{amsmath}
\usepackage{amsfonts,amssymb}
\usepackage{mathrsfs}

\newtheorem{theorem}{Theorem}[section]
\newtheorem{lemma}[theorem]{Lemma}
\newtheorem{remark}[theorem]{Remark}
\newtheorem{definition}[theorem]{Definition}
\newtheorem{proposition}[theorem]{Proposition}
\newtheorem{corollary}[theorem]{Corollary}

\numberwithin{equation}{section}
\newcommand{\LOP}{L^2(\Omega)}
\newcommand{\f}{\varphi}
\newcommand{\Rset}{\mathbb{R}}
\newcommand{\R}{\mathbb{R}}
\newcommand{\Nset}{\mathbb{N}}
\newcommand{\Zset}{\mathbb{Z}}
\newcommand{\C}{\mathbb{C}}

\newcommand{\Toas}{\mathop{\longrightarrow} \limits^{a.s.}}
\newcommand{\spM}{M} 
\newcommand{\spP}{\Pi} 
\newcommand{\spPt}{{\Omega}} 
\newcommand{\p}{\mathbf{p}} 
\newcommand{\PP}{\bs{\pi}} 
\newcommand{\PPt}{\mathbb{P}} 
\newcommand{\mi}{m} 
\newcommand{\mat}{{\mathcal B}} 
\newcommand{\eps}{\varepsilon}
\newcommand{\epsef}{\varepsilon^{\mathrm{eff}}}
\newcommand{\bs}{\boldsymbol} 
\newcommand{\muef}{\mu^{\mathrm{eff}}}

\newcommand{\E}{\mathbb{E}}
\newcommand{\disp}{\displaystyle}

\renewcommand{\O}{\tilde\Omega}
\renewcommand{\o}{\tilde\omega}
\newcommand{\Div}{\mathrm {div}\,} 
\newcommand{\Divx}{\mathrm {div_x}\,} 
\newcommand{\Divs}{\mathrm {div^s}\,} 
\newcommand{\Grads}{\mathrm {\nabla^s}\,} 
\newcommand{\DS}{\displaystyle}

\newcommand{\beps}{\boldsymbol{\varepsilon}}
\newcommand{\by}{\boldsymbol{y}}
\newcommand{\bepsilon}{\boldsymbol{\varepsilon}}
\newcommand{\brho}{\boldsymbol{\rho}}
\newcommand{\btheta}{\boldsymbol{\theta}}
\newcommand{\bsigma}{\boldsymbol{\sigma}_0}
\newcommand{\bpsi}{\boldsymbol{\psi}}
\newcommand{\dis}{\displaystyle}
\newcommand{\tor}{Y} 
\newcommand{\med}{\medskip\noindent}

\newcommand{\loc}{\mathrm {loc}}
\newcommand{\dist}{\mathrm {dist}\,}
\newcommand{\DomD}{\mathcal D}

\newcommand{\ov}{\overline}
\newcommand{\wto}{\rightharpoonup}
\newcommand{\twoscale}{\makebox[0pt][l]{$\rightharpoonup$}\rightharpoonup}

\newcommand{\stwoscale}{\makebox[0pt][l]{$\rightarrow$}\rightarrow}

\newcommand{\QED}{\hfill\ensuremath{\square}}

\title{Resonant effects in random dielectric structures
}
\date{}
\author{Guy Bouchitt\'e, Christophe Bourel, Luigi Manca}
\begin{document}
\title{Resonant effects in random dielectric structures}
\address{Guy BOUCHITTE: IMATH, Universit\'e du Sud Toulon-Var, 83957 La Garde cedex, France}
\address {Christophe BOUREL: LMPA, Universit\'e du littoral c\^ote d'Opale, 62228 Calais Cedex, France}
\address{Luigi MANCA:  LAMA, Universit\'e de Marne La Vall\'ee, 77454 Marne la Vall\'ee Cedex 2, France}
%
%
%
%

\maketitle
\begin{abstract}

In \cite{BF,prl,njp}, a theory for artificial magnetism in two-dimensional
photonic crystals has been developed for large wavelength using homogenization techniques.
In this paper we pursue this approach within a rigorous stochastic framework:
dielectric parallel nanorods are randomly disposed, each of them having, up to a large scaling factor, a random permittivity $\eps(\omega)$ whose law is represented by a density on a window $ \Delta_h=[a^-,a^+]\times [0,h]$
of the complex plane.
We give precise conditions on the initial probability law (permittivity, radius and position of the rods) under which the homogenization process can be performed leading to a deterministic dispersion law for the effective permeability
with possibly negative real part. Subsequently a limit analysis $h\to 0$, accounting a density law of $\eps$ which concentrates on the real axis, reveals singular behavior due to the presence of
resonances in the microstructure.
\\

%

\end{abstract}
\section{Introduction}

\paragraph{Physical background} \ The paper is motivated by recents studies on metamaterials which are artificial structures that can be described,
 in certain ranges of frequencies, by homogeneous parameters. These materials enjoy very unusual properties not encountered in natural materials. In particular, it is possible to design devices in such a way that their effective permittivity and permeability be both negative, resulting in an
effective negative index, as studied years ago by Veselago \cite{Veselago}. Several geometries have
been proposed, notably the split ring resonators structure, that can produce artificial
magnetism \cite{bou_schw,kohn_ship,pendry2}. Such a structure can be combined  with a  metallic wire mesh structure
 in order to produce meanwhile a negative permittivity
\cite{pendry,Shelby,Smith,Houck,boufel_random_media,Felbacq94,cicp}. In ref. \cite{Pendry_mu}, O'Brien and Pendry also suggested that it is
possible to obtain an artificial magnetic activity in photonic crystals made of dielectric
rods with a strong permittivity. In that case, there are internal resonances (Mie resonances) in the rods at wavelengths which are large compared with the period of the structure. Around
the resonances, the magnetic  field is strongly localized inside the rods: this produces
a loop of displacement current inducing a microscopic magnetic moment. All the
microscopic moments add up to a collective macroscopic moment, resulting in artificial
magnetism. In \cite{prl,njp} a rigourous justification for this phenomenum 
 was proposed in the particular case of a  photonic crystal made
of infinitely long parallel rods illuminated by 
 a transverse magnetic  polarized incident field: by using a two-scale  analysis 
and by exploiting the spectral decomposition of the the 2D-Dirichlet-Laplace operator on 
the section of the rods, it was possible to establish a frequency dependent effective permeability law with sign changing real part.
   Next in \cite{njp}, the influence of random variations in the parameters of the rods  has been studied numerically namely the position of the rods and their permittivity.
 For the well posedness of the diffraction problem, a small dissipation (positive imaginary part)
 has been assumed for the permittivity of the random rods and, as expected, the influence of this dissipation combined with disorder tends to attenuate the resonance effects.

In this paper we will study the previous dielectric rod structure in a full rigorous random framework.
The incident magnetic field is assumed to be polarized in the vertical direction $e_3$ and parallel to the rods.  By invariance in the $x_3$- direction, the diffration problem can be 
therefore reduced in a 2D framework. The generalization of our results to a full 3D-framework is much more involved and
could be foreseen as a challenging perspective (see \cite{bou_bou_cras} in the  periodic deterministic case).

\paragraph{Description of the random obstacle and scaling}\quad 
We consider a bounded domain $\mat\subset \Rset^2$ in which 
 small  circular inclusions are randomly disposed. 
 These inclusions occupy a subregion $\DomD_\eta(\omega)\subset \mat$
  representing the sections of the highly dielectric rods. 
  Here  $\eta$ is an infinitesimal quantity accounting the distance between inclusions while $\omega$ represents the random dependence \textit{in contrast with the standard notation for the angular frequency commonly used in waves theory}.
   
\begin{figure}[!ht]
\begin{center}
\includegraphics[width=0.75\linewidth]{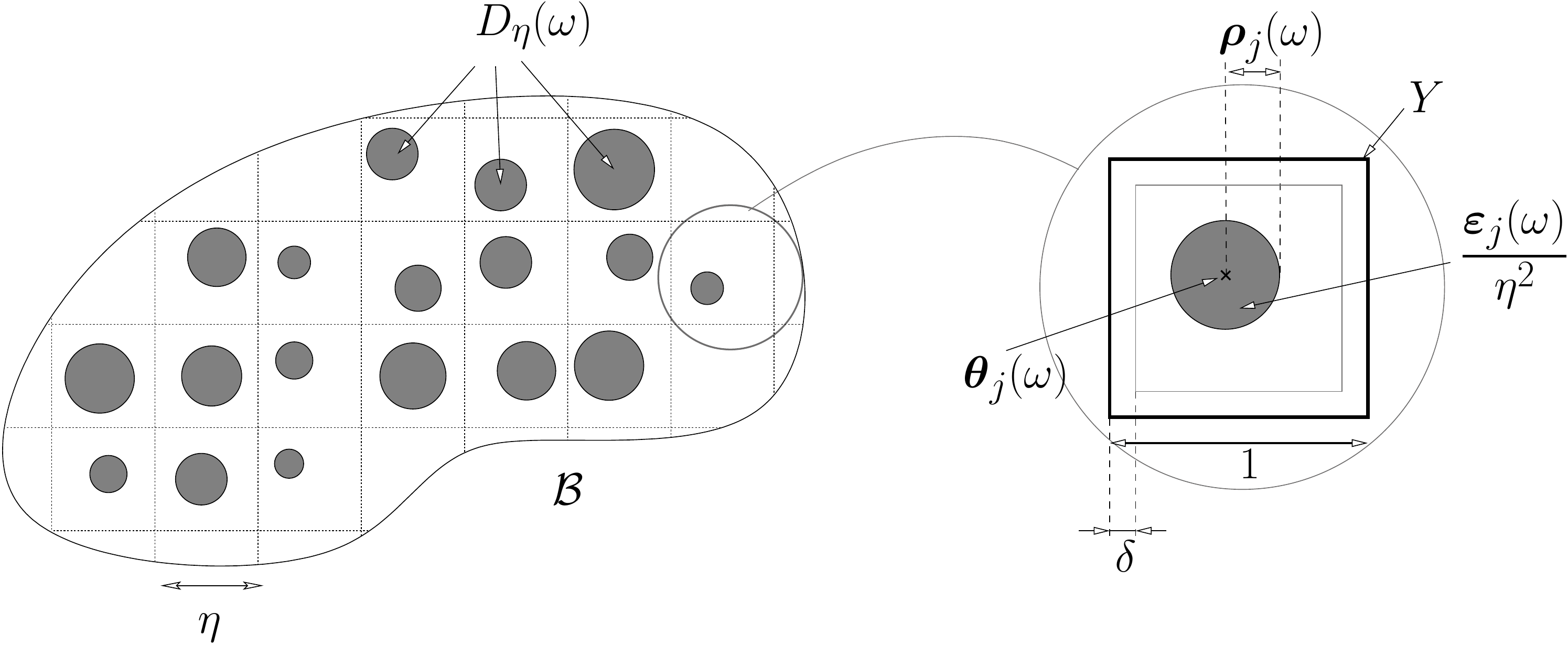}
\caption{Description of the random obstacle \label{fig:dom}}
\end{center}
\end{figure}

  \med
 More precisely, to every event $\omega$  (in a probability space $\Omega$ defined later),
 we associate a sequence of disks  $\{B(\boldsymbol{\theta}_j(\omega),\boldsymbol{\rho}_j(\omega));\ j\in \Zset^2\}$  whose centers $\boldsymbol{\theta}_j(\omega) $ and radius $\boldsymbol{\rho}_j(\omega) $ are  chosen randomly independent
 and so that $B(\boldsymbol{\theta}_j(\omega),\boldsymbol{\rho}_j(\omega)) \subset\subset Y:=(0,1)^2$. Then, letting $\bs{y}(\omega)\in Y$ be a uniformly distributed random translation of the unit lattice of $\Rset^2$, we define
 the following random subset of $\R^2$:
\begin{equation}\label{def:Domega}
\DomD(\omega):=\bigcup_{j\in \Zset^2}\DomD^j(\omega),\qquad 
\DomD^j(\omega):=j-\bs{y}(\omega)+B\big(\boldsymbol{\theta}_j(\omega),\boldsymbol{\rho}_j(\omega)\big)
\end{equation}
Then after a $\eta$ dilatation, we determine a fine perforated domain in $\mat$ by setting
\begin{equation}
\label{def:Deta}
\DomD_\eta(\omega):=\bigcup_{j\in J_\eta(\omega)}\DomD^j_\eta(\omega),\qquad \DomD^j_\eta(\omega):=\eta\DomD^j(\omega)
\end{equation}
$$ \text{where}\qquad J_\eta(\omega)=\{ j\in\Zset^2\ |\ \eta(j-\bs{y}(\omega)+\tor)\subset \mat\}$$
%
For technical reasons  (see Remark \ref{r.delta}), we also assume that there exits $\delta\in (0,\frac12)$ such that $\dist(\btheta_j(\omega),\partial \tor)>\rho_j+\delta$ for all $j\in \Zset^2$ and all $\omega\in\Omega$.
The scattering obstacle $\DomD_\eta(\omega)$ is depicted in Figure \ref{fig:dom}.

A fundamental issue is now to choose a suitable scaling for the permittivity contrast between the rods and the surrounding matrix. In order that resonances occur for an incident wavelength which is large with respect to the radius of the rods (of order $\eta$), we need that the relative permittivity in them to be of order $\eta^{-2}$. We refer for instance to  
 \cite{Pendry_mu,prl,njp} for further comments on this choice.
 To simplify we will assume that the permittivity in the matrix $\mat\setminus\DomD_\eta(\omega)$ is the same as in the vacuum.
 In contrast in each rod, namely on $\DomD^j_\eta(\omega)$, we set the relative permittivity to be  $\eta^{-2}\, {\boldsymbol{\varepsilon}_j(\omega)}$ where $\{{\boldsymbol{\varepsilon}_j(\omega)}, j\in J_\eta(\omega)\}$ 
 are mutually independent equi-distributed random variables ranging in the set  $\C^+$ of complex numbers with positive imaginary part.
 From the physical point of view, the real numbers  $\brho_j (\omega)\sqrt{ \Re (\beps_j(\omega)) }$  correspond to the``optical radius'' of the dielectric rods. The choice of the scaling in $\eta^{-2}$
  while the radius of the rods is of order $\eta$ allows to keep this optical
radius of order one. 
 
\paragraph{Diffraction problem}

The set of random rods is illuminated by a monochromatic incident wave traveling in the $H||$ mode.
The incident magnetic field is therefore of the form $H^{\rm inc}(x,t)=u^{\rm inc}(x_1,x_2)e^{-i\, \nu t}\, e_3$ where $\nu$ denotes the angular frequency and $u^{\rm inc}$ solves Helmholtz equation $\Delta u^{\rm inc} + k_0^2 u^{\rm inc}=0$ in $\Rset^2$ where $k_0:=\nu /c= 2\pi / \lambda$ is  the wave number ($c$ is the light speed
in the vacuum and $\lambda$ the wavelength).

Here and in what follows, in order to simplify the mathematical framework, the unit of length has been normalized so that
roughly $\lambda$ represents the ratio between the physical wavelength and the size of the obstacle $\mat$.
Therefore our normalized $k_0$ and $ \lambda$ are \textit{dimensionless quantities}.

Then as usual the total electromagnetic field solving Maxwell's system is completely
determined by the magnetic field  $H_\eta(x,t)=u_\eta(x_1,x_2)e^{-i\, \nu t}\, e_3$ 
where $u_\eta$ solves   
\begin{equation}\label{eq.eta}
\Div \left(a_\eta(x,\omega)\nabla u_\eta(x,\omega)\right)+k_0^2\, u_\eta(x,\omega)=0\quad x\in \Rset^2,\, \omega\in \Omega
\end{equation}
Here the complex function $a_\eta(x,\omega)$ represents the inverse of the local permittivity.
In view of the scaling adopted before, it is given by
\begin{equation}\label{eq.a_eta_ini}
a_\eta(x,\omega)=1_{\Rset^2\setminus\DomD_\eta(\omega)}(x)+\sum_{j\in J_\eta(\omega)}\frac{\eta^2}{\boldsymbol{\varepsilon}_j(\omega)}1_{\DomD^j_\eta(\omega)}(x).
\end{equation}
In addition the \textit{diffracted} field $u_\eta^d=u_\eta-u^{\rm inc}$ satisfies the outgoing Sommerfeld radiation condition
\begin{equation}\label{SM}
\lim_{r\to \infty} \sqrt{r} \, \left(\partial u_\eta^d/\partial r - ik_0 u_\eta^d\right) \ =\ 0.
\end{equation}
  
  Let us emphasize that the randomness of the system and all the physical informations of the model are determined by the probability law $\p$ on $ \tor\times[0,1/2]\times \C^+ $ shared by the random triples $({\boldsymbol{\theta}_j(\omega)}, {\boldsymbol{\rho}_j(\omega)},{\boldsymbol{\varepsilon}_j(\omega)})$. 
 We point out also  that in this model {\em no magnetic activity} is present in the rods 
 namely their permeability is taken to be the same as in the vacuum. 
  
  \paragraph{Asymptotic issues} We are interested in the behavior of diffraction problem \eqref{eq.eta} for infinitesimal values of parameter $\eta$.
  In particular we want to show that, under suitable conditions on the probability law $\p$,  the solution $u_\eta(x,\omega)$ does converge almost surely to a {\em  deterministic} limit. We expect this limit to be characterized as the unique solution of a new diffraction problem in which the obstacle $\mat$ is filled with an homogeneous material  characterized by suitable effective permeability
and permittivity tensors.

\med
 Our goal is twofold: 

- \ First we want to derive  conditions on the given probability law $\p$ under which uniform $L^2$ -estimates hold for the sequence $\{u_\eta\}$ allowing to undergo the homogenization process
in a similar way as in the deterministic case studied in \cite{BF}.

-\ Second we wish to study the homogenized dispersion law for the effective permittivity
in term of $\p$. In particular an important issue is the influence of the disorder when the support of the probability law of the ${\boldsymbol{\varepsilon}_j(\omega)}$'s concentrates along the real axis. Surprisingly the limit anaysis will reveal that the homogenized permeability law
 retains a non vanishing dissipative part.

\paragraph{Outline}
The paper is organized as follows.  The main homogenization result is stated in Section 
\ref{s.mainresult} after introducing suitable notations and definitions.
In Section \ref{law}, we analyze the effective permittivity law emphasizing on the
combined role of resonances and randomness. In particular, we discuss the behavior of the limit field with respect to some critical choice of the law $\p$.
Then after a detailed review of the stochastic framework in Section 4,  we present in Section 5 the two-scale analysis of the diffraction problem.
 The Section 6 is devoted to the proof of the main result (Theorem \ref{main})  followed by a technical Appendix.

%
%
%



%

\section{The main homogenization result} \label{s.mainresult}

\paragraph{Notations} 
 We denote by $\{e_1, e_2\}$ the canonical base of $\R^2$.
 Sometimes $e_1, e_2$ will be identified as elements of $\Zset^2$. 
 Given $x,y \in\R^2$, $|x|$ denotes the Euclidean norm and $x\cdot y$ the scalar product.
 Accordingly, for every $r>0$, $B(x,r)$ is the open ball centered in $x$ of radius $r$ and  we denote $B_r :=B(0,r)=\{ |x|< r\}$.
  For every open set $A\subset\R^2$, $C^\infty_c(A)$ stands for the space of $C^\infty$ complex valued functions whose support
 is a compact subset of $A$.
 Given a complex number $z\in\C$, we denote by $\Re(z)$ and $\Im(z)$ its real and imaginary  parts.
 We denote $\C^+:=\{z\in \C\ :\ \Im(z)\ge 0\}$. Given $\delta$ a fixed parameter such that $0<\delta<1/2$, we define
\begin{equation}\label{eq.def_Omeg}
 \spM=\{ (\theta,\rho, \varepsilon)\in \tor\times[0,1/2]\times \C^+: \dist(\theta, \partial\tor)\geq \rho+\delta\}\, .
\end{equation}
Eventually we will consider $Y :=(0,1)^2$ as the unit periodicity cell in $\R^2$.
Accordingly we denote by $W^{1,2}_\sharp(Y)$  the set of restrictions to $Y$ of functions in $W^{1,2}_{\rm loc}(\R^2)$ 
  which are $Y$-periodic. This space is naturally endowed with a Hilbert space structure.
\paragraph{Probability space}
A general event $\omega\in\Omega$ will be represented as
\[
\omega=\big((\theta_j,\rho_j,\varepsilon_j)_{j\in\Zset^2},y\big)= \big(m,y\big),
\]
where $m=(\mi_j)_{j\in\Zset^2}$ is a sequence of elements in $M$ and $y\in Y$.
We consider a reference  Borel probability law $\p$ on $M$ which, for the sake of simplicity, we assume to be \textit{compactly supported}. 
Then our general probability space $(\Omega,\mathcal{A}, \PPt)$ reads 
\[
\Omega:=\spP\times Y\quad,\quad \mathcal{A}=\mathcal{F}\otimes\mathscr{B}(Y)\quad,\quad \PPt=\PP\otimes \mathcal L^2\ ,
\]
being $\mathcal L^2$ the Lebesgue measure and the space $\spP:=\spM^{\Zset^2}$  (metrisable
and compact) endowed with the product measure $\PP=\bigotimes_{\Zset^2}\p$  and  Borel tribe   $\mathcal{F}=\bigotimes_{\Zset^2}\mathscr{B}(\spM)$.

\med
In the following $C(\Omega)$ will denote the space of {\em bounded continous} functions on $\Omega$.
The expectation of a random function $f(\omega)$ in $L^1(\Omega,\PPt, \mathcal{A})$ is denoted by the symbol $\E(f)$.

\med
We indicate in bold symbol $\btheta_j(\omega)$ (respectively
$\brho_j(\omega)$, $\beps_j(\omega),\by(\omega)$)   the random variable  associated with the projection $\omega\in\Omega \mapsto \theta_j$  (resp. $\omega\mapsto \rho_j, \eps_j, y$).
The triples $(\btheta_j(\omega), \brho_j(\omega),\beps_j(\omega))$ are independent and identically distributed random variables,  which seen as functions on $M$ 
share the given probability law $\p$.
Accordingly the random  disk $\DomD_\eta^j(\omega)$  is given by (\ref{def:Deta}).
\medskip

Next we introduce the following Borel subsets of $\Omega$:
\begin{equation}\label{Sigma}
\Sigma=\{ \omega\in \spPt: |y-\theta_0|< \rho_0\},\qquad \Sigma^*=\spPt\setminus \ov{\Sigma}.
\end{equation}
We notice that the real number  $\PPt(\Sigma)\in (0,1)$ represents the average \textit{filling ratio} of the inclusions in $\mat$. Indeed, recalling that the last marginal of $\PPt$  coincides with the Lebesgue measure on $Y$,  by the law of large numbers it holds almost surely:
$$\lim_{\eta\to 0}  \frac{|\DomD_\eta(\omega)|}{|\mat|}\ =\ \PPt(\Sigma)\ =\ \pi \int_M \rho^2 \, \p(d\theta d\rho d\eps)\ . $$

\begin{remark} \label{r.delta}\upshape
If $(\theta,\rho,\varepsilon)$ belongs to $\spM$ then the disk $B(\theta,\rho)$  is contained into $\tor$ and its distance to the boundary $\partial \tor$ is at least $\delta$.
The prescription of such constant $\delta>0$ will be crucial in order to obtain a \textit{uniform}  $L^2(\mat)$ estimate for $\{u_\eta\}$ for $\omega$ running over $\Omega$
(see in particular Lemma \ref{l.laplace}).
\end{remark}


\paragraph{Effective permittivity} The effective permittivity tensor $\eps^{\mathrm{eff}}$ 
that  arises from our asymptotic analysis  turns out to be symmetric positive and \textit{ independent} on the frequency.
It relies only on the geometry of the rods, namely on the distribution law of their radius and centers. 
In fact the computation of $\eps^{\mathrm{eff}}$ falls in the  homogenization theory for Neumann problems on random perforated sets 
as developed for instance in  \cite{JKO}. We have the following variational characterization, for every $z\in \R^2$
\begin{equation}\label{eq.epsef}
\epsef z \cdot z \ :=\ \inf \left\{   \int_\Omega |\sigma|^2 \, d\PPt \ :\ \sigma \in L^2_{\rm sol} (\Omega; \R^2)
\ ,\ \sigma =0 \ \text{ in $\Sigma$} \ ,\ \E(\sigma) =z \right\} \,
\end{equation}
where $ L^2_{\rm sol} (\Omega; \R^2)$ denotes the set of solenoid random vector fields as defined in \eqref{solenoid}.
The minimum above is taken over a closed subspace of $(\LOP)^2$ which roughly corresponds to vector fields 
on $\R^2$ which are divergence free and vanish on the random set $\DomD(\omega)$ (see Proposition \ref{p.diff_stoc}).
It is easy to check that, for very $z=(z_1,z_2)\in\R^2$, the  solution $\sigma$ to \eqref{eq.epsef} exist and is unique. By linearity,
it is given by
 $ \sigma = z_1\, \sigma^1 + z_2\, \sigma^2$ where, for $i=1,2$,  $\sigma^i$ denotes the solution associated with $z=e_i$.
 Therefore, the effective tensor $\eps^{\mathrm{eff}}$ can be witten as follows
 \begin{equation}\label{epstensor}
\eps^{\mathrm{eff}}_{i,j} \ =\ \E( \sigma^i \cdot \sigma^j)\ .
\end{equation}
Let us mention some straightforward estimates for that tensor. Firstly
we observe that, if we set $ Y_\delta:=\{y\in Y\ :\ {\rm dist}(y, \partial Y) >\delta\}$, then  by \eqref{eq.def_Omeg} and \eqref{Sigma} it holds $\Sigma\subset \Pi \times Y_\delta$ and therefore  vector fields $\sigma= \sigma(y)$ are admissible if they belong to the class $\mathcal{S}_\delta$ of restrictions to $Y$ of $Y$-periodic, divergence free fields  of $L^2_{\rm loc} (\R^2)$ vanishing in $Y_\delta$ (see $ii)$ of Proposition \ref{p.diff_stoc}).
Therefore the infimum in \eqref{eq.epsef} is finite and attained in a unique $\sigma$. Futhermore the following upper bound holds:
\begin{equation}
 \epsef  \ \le \ C(\delta) \, {\rm I}_2 \quad {\rm with} \quad
C(\delta) :=  \inf \left\{   \int_Y |\sigma|^2  \, dy \ :\  \sigma \in \mathcal{S}_\delta \ ,\ \int_Y \sigma(y) = \bs{e}\right\}\ , 
\label{maxepsef}
\end{equation}
where $\bs{e}$  is an arbitrary unit vector and ${\rm I}_2$ denotes the two by two identity matrix.
Note that the constant $C(\delta)$ represents the inverse of the effective diffusion coefficient in the (determinist) case of a 
periodic perforated domains with square holes (it blows up to infinity as $\delta \to 0$).
On the other hand the effective tensor $\epsef$ is non degenerate. In fact as can be easily checked by Young inequality, we have the following lower bound
\begin{equation}
\label{minepsef}
\epsef  \ \ge\ \frac1{\PPt(\Sigma^*)} \ {\rm I}_2 
\end{equation}
Indeed one has $ \int_{\Sigma^*} |\sigma|^2  \ge 2 z\cdot z^* - |z^*|^2  \, \PPt(\Sigma^*)$,  for every admissible $\sigma$ and $z^*\in \R^2$, thus \eqref{minepsef} by taking the supremum in $z^*$.

\med
%
%
\begin{remark} \upshape{Unfortunately the formula \eqref{eq.epsef} is far to be explicit and $\epsef$ is not necessarily a scalar tensor
unless the probability distribution of the center $\btheta_0(\omega)$ is invariant by $\pi/2$ rotations.
Note that if the radius of the rods is a constant $\rho_0$ (i.e. the probability law of $\brho_0$ is a Dirac mass)
then $\epsef$ is scalar , does not depend on the law of the center $\btheta_0$ and  \eqref{eq.epsef} agrees with the 
classical formula of the deterministic case (see \cite{Cio}). 
}\end{remark} 

\paragraph{Effective permeability} The effective permeability $\mu^{\mathrm{eff}}$ depends on the frequency $\nu$ and its behaviour is ruled by the spectral values of the Dirichlet Laplace operator in the two dimensional unit disk $D:=B(0,1)$.
Let us denote by $\bsigma:=\{\lambda_n,n\in\Nset^*\}$ with $0<\lambda_1 < \lambda_2 \le  \cdots \le\lambda_n\le \cdots$ the eigenvalues of this operator and let $\{\varphi_n,n\in\Nset^*\}$ be an associated orthonormal basis of eigenvectors in $L^2(D)$. 
We set 
$$c_n :=\int_D \varphi_n \qquad  \text{(thus $1= \sum c_n \varphi_n$ in $L^2(D)$ and 
  $\sum |c_n|^2 = \pi$).
} $$
 Then, recalling the definition of $\Sigma$ in \eqref{Sigma},  we introduce the random variable $\Lambda(\omega)$ 
\begin{equation} \label{eq.lambda}
\Lambda(\omega)=
\begin{cases}
\dis 1+\sum_{n=1}^{+\infty}
c_n\,\frac{k_0^2\beps_0(\omega)\brho_0^2(\omega)}{\lambda_n-k_0^2\beps_0(\omega)\brho_0^2(\omega)}
\, \varphi_n\Big(\frac{\boldsymbol{y}(\omega)-\btheta_0(\omega)}{\brho_0(\omega)}\Big)&
\mbox{if }\omega\in \Sigma\\
1&\mbox{if }\omega\in \Sigma^*\ .
\end{cases}
\end{equation}

The effective permeability $\mu^{\mathrm{eff}}(k_0)$  can be then 
expressed as a function of the incident wave number $k_0$ (recall $k_0= \frac{\nu}{c}$)
as follows:
\begin{equation} \label{def.mueff}
\mu^{\mathrm{eff}}(k_0)=\mathbb E\big(\Lambda(\omega)\big)=1+\sum_{n=1}^{+\infty} c_n^2\int_M\frac{\varepsilon\rho^4k_0^2}{\lambda_n-\varepsilon\rho^2k_0^2}\,{\rm d}\p(\theta,\rho,\eps).
\end{equation}
%
We notice that the random fluctuations of the center $\theta$ of the disks have no effect 
on the value of $\muef$ (which depends only on the $(\rho,\eps)$-marginal of $\p$).

\paragraph{Homogenization result} Recalling (\ref{eq.epsef}) and (\ref{def.mueff}), we  introduce  the local permittance and permeability functions over all $\Rset^2$ as follows:
\begin{equation}\label{globalepsmu}
\begin{array}{lll}
\mathbf{A}(x) &:= & 1_{\mat}(x) \, ({\epsef})^{-1}\ +\  1_{\Rset^2\setminus\mat}(x)\ I_2 ,  \\
\mathbf{b}(x) &:= &  1_{\mat}(x)\, \mu^{\mathrm{eff}}(k_0) \ +\ 1_{\Rset^2\setminus\mat}(x)\ .
\end{array}
\end{equation}
The limit diffraction problem  takes then the form
\begin{equation} \label{e.limiteq}
\begin{cases} 
\mathrm{div}\left(\mathbf{A}(x)\nabla u\right)\, +\, k_0^2\, \mathbf{b}\, u=0,\\ 
u-u^i\quad\text{satisfies the outgoing radiation condition (\ref{SM})}
\end{cases}
\end{equation}
Here $u\in W^{1,2}_{\rm loc}(\R^2)$ represents the ``effective'' magnetic field. As will be seen later (see Theorem \ref{main}), it is defined 
in such a way that it agrees with the pointwise limit of $u_\eta$ outside the obstacle \textit{but it differs from the weak limit of $u_\eta$ inside}. 

Let us emphasize that the first equation in \eqref{e.limiteq} has to be understood in the distributional sense.
In particular the fact that $\nabla u,\ \mathrm{div}\left(\mathbf{A}(x)\nabla u\right)$ belong to $L^2_{\loc}$ imply the following 
transmission conditions on the boundary of $\mat$:
\begin{equation}\label{transmission}
  u^+ = u^-   \quad,\quad \partial_n u^+ =  \left(({\epsef})^{-1}  \nabla u^-\right) \cdot n\ ,
\end{equation}
where superscript $\pm$ indicate the traces from outside and inside for $u$ and the normal derivative $\partial_n u$
(being $n$ the exterior normal on $\partial\mat$).
These transmission conditions indicate that  $u$ (as we have defined inside the obstacle) represents the magnetic field seen from the exterior.

\med
The main result of this paper stated below requires two hypotheses on the probability law $\p$:
  
\begin{equation} \label{eq.hyp}
 \exists \mbox{ $r>0$ such that\qquad} \int_\spM \left[\dist \big(\varepsilon\rho^2  k_0^2,\bsigma\big)\right]^{-(2+r)} d\p<\infty \ ,
\end{equation}

 \begin{equation} \label{eq.hyp_im}
       \p\Big(\{(\theta,\rho,\eps)\in M \ |\ \Im m(\eps)>0\}\Big)>0\ .
      \end{equation}


\bigskip
\begin{theorem}\label{main}
 Assume that (\ref{eq.hyp}) and (\ref{eq.hyp_im}) hold.
Then the  diffraction problem (\ref{e.limiteq}) admits a unique  solution $u$ and, for $\PPt$-almost all $\tilde\omega\in \Omega$, there holds
\[
 u_\eta(\cdot,\tilde\omega) \ \wto \ \mathbf {b}\, u \quad\text{weakly in $L^2_{\mathrm{\loc}}(\Rset^2)$}\ .
\]
In addition, on compact subsets of $\Rset^2 \setminus\mat$ (where $\mathbf {b}=1$),  the  convergence   $u_\eta(\cdot,\tilde\omega)\to u$
is uniform as well as for all derivatives.  

\med
Eventually the defect of strong convergence in $L^2(\mat)$ can be recast  by the following equalities 
\begin{equation} \label{e.stron_conv_intro}
\lim_{\eta\to0} \int_\mat |u_\eta(x,\tilde\omega)|^2\,dx\ =\ \mathbb E\left(\int_\mat|u_0(x,\cdot)|^2\,dx\right)
\ =\ \ \mathbb E(|\Lambda(\cdot)|^2)\, \int_\mat| \mathbf{b}(x)\,  u(x)|^2\,dx
\end{equation}
being $\Lambda(\omega)$ given by (\ref{eq.lambda}) and $
u_0(x,\omega) :=\,  u(x)\big(\Lambda(\omega)\,1_{\mat}(x)+1_{\Rset^2\setminus\mat}(x)\big)$.
 \end{theorem}
%
%

\begin{remark}{\upshape
This result generalizes to the random setting the results of \cite{BF}. More precisely, the periodic case is recovered by taking the probability $\p$ to be a Dirac mass at $(\btheta_0,\brho_0,\beps_0)$. Conditions (\ref{eq.hyp}) and (\ref{eq.hyp_im}) are trivially satisfied provided
 $\Im(\beps_0) >0$. In this case the probability space $(\Omega,\PPt)$ can be identified with $(Y,\mathcal{L}^2)$ and the function $u_0=u_0(x,y)\in L^2_{\rm loc}(\R^2;W^{1,2}_\sharp(Y))$ coincides with the classical two-scale limit of $u_\eta$ (see \cite{Allaire}). Furthermore, the convergence in (\ref{e.stron_conv_intro}) yields the \textit{ strong} two-scale convergence of 
$u_\eta$ to $u_0(x,y)$ which is an improvement of the result presented in \cite{BF}.
In fact this improvement of the convergence can be extended
to the random by using the dynamical system approach
developed in Section 4 and the map $T$ defined in (\ref{mapT}). More precisely we will show that
 for any ball $B_R$, large enough,  there holds for almost all $\o\in \O$
\begin{equation}
\lim_{\eta\to0^+}\int_{B_R}\left| u_\eta(x,\tilde\omega) - u_0(x,T_{\frac{x}{\eta}}\tilde\omega)\right|^2dx =0\ .
\end{equation}

}
\end{remark}

\begin{remark} \upshape{
We can see easily that the imaginary part of $\mu^{\mathrm{eff}}(k_0) $ is positive if and only if the condition
\eqref{eq.hyp_im} is
verified. Thanks to this assumption, the limit problem (\ref{e.limiteq}) turns out to be well posed. Notice that (\ref{eq.hyp}) and (\ref{eq.hyp_im}) are trivially satisfied if the support of the permittivity  law in the rods  does not intersect the real axis.  If it is not the case, it looks natural to add a small dissipative parameter so that our homogenization result applies and then to send this parameter to zero.
This important issue is discussed in Section \ref{law}.   
}\end{remark}

%


\section{Frequency dependent effective permeability law}\label{law}



In the context of finding new metamaterials with magnetic activity,
the homogenization result in Theorem \ref{main} demonstrates that random dielectric structures are theoretically able to produce  permeability laws  like in (\ref{def.mueff}). In particular effective permeability coefficients $\muef$ with
 large negative real part are possible in suitable ranges of frequencies. At this stage, it is worth to study the behaviour of the effective permeability $\muef(k_0)$ with repect to the probability law $\p$ characterizing the random rods.

\subsection{Resonant effects} Let us rewrite (\ref{def.mueff}) as
$$
\dis \mu^{\mathrm{eff}}(k_0)\ =\ 1+\sum_{n\in I} c_n^2\int_M\frac{\rho^2 k_0^2}{k_n^2(\rho,\eps)-k_0^2}\,{\rm d}\p(\theta,\rho,\eps)\ ,
$$
being $I:=\{n\ge 1\ :\ c_n\not=0\}$ and $k_n:=k_n(\rho,\eps) \in \C^+$  such that $\dis k_n^2=\frac{\lambda_n}{\eps\, \rho^2}.$

The internal resonant frequencies are characterized by the real random numbers 
$\{\nu_n, n\in I\}$ where $\nu_n=  \sqrt{\Re e(k_n^2)}$.
Indeed, if the random variable $\Im ( \eps)$ remains small, a huge variation of $\muef$ will appear when $k_0$ passes through one of the $\nu_n$'s with  a change of sign of the real part.
  On the other hand it is clear that the random fluctuations of 
 the $\nu_n$'s  contribute to damp the leading term in the power series.
 
  To illustrate these facts,  we have considered rods with constant relative permittivity $\eps_r=100+5i$ and compared the
 situations where the reference radius $\rho_0=0.375$  is kept constant 
or is uniformly distributed on interval $I:=[0.3, 0.45]$.
The first case is  purely determinist  and corresponds to taking $\p$ to be a Dirac mass at $(\rho_0, \theta_0, \eps_r)$
(the position of center $\theta_0$ has no influence) whereas in the second case we have $\p = 1_{\rm I}
(\rho) d\rho\otimes \delta(\theta-\theta_0)\otimes \delta(\eps-\eps_r)$.

\begin{figure}[!h]
\begin{minipage}[b]{0.47\linewidth}
\includegraphics[width=\linewidth]{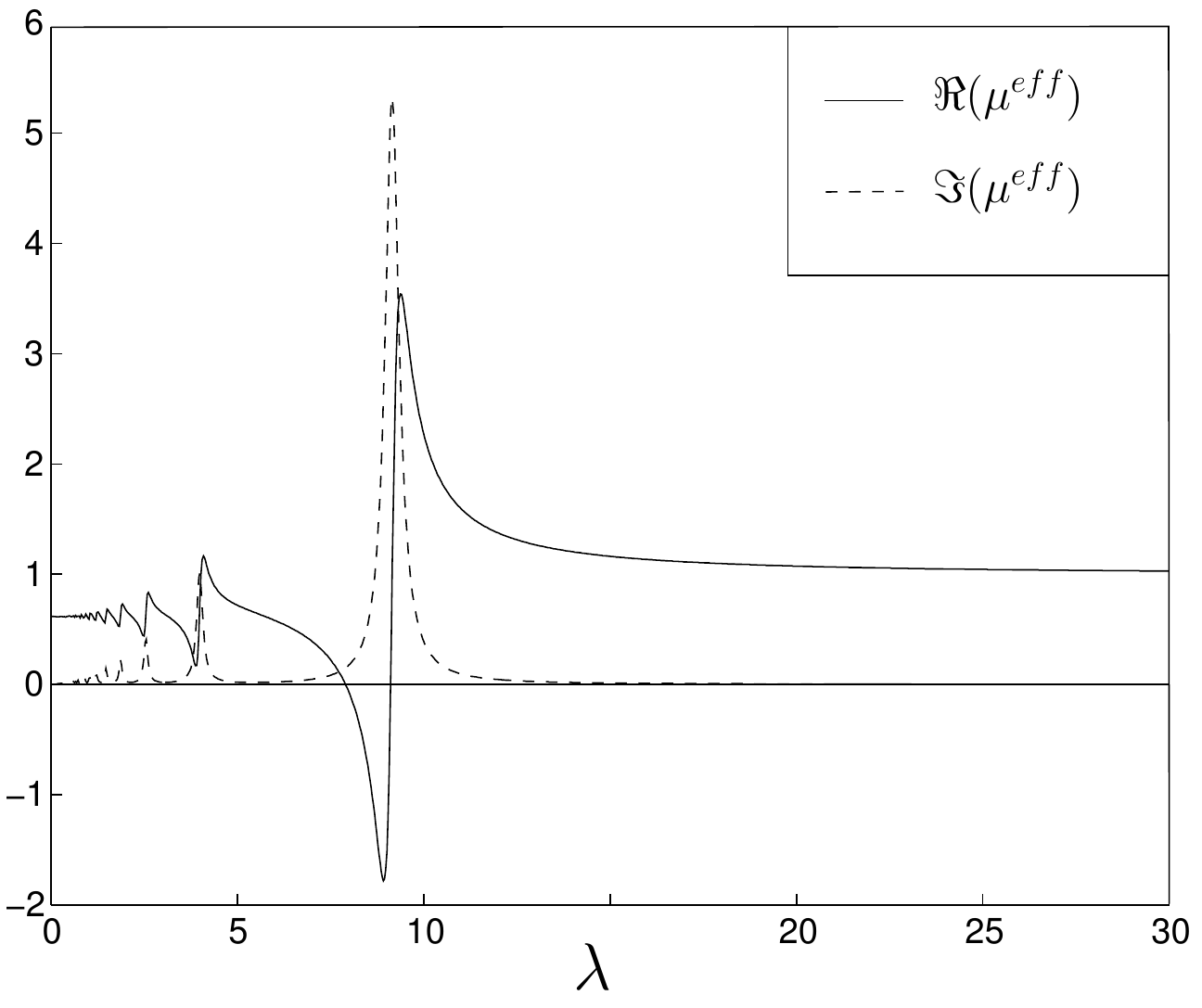}
\end{minipage}
\begin{minipage}[b]{0.48\linewidth}
\includegraphics[width=\linewidth]{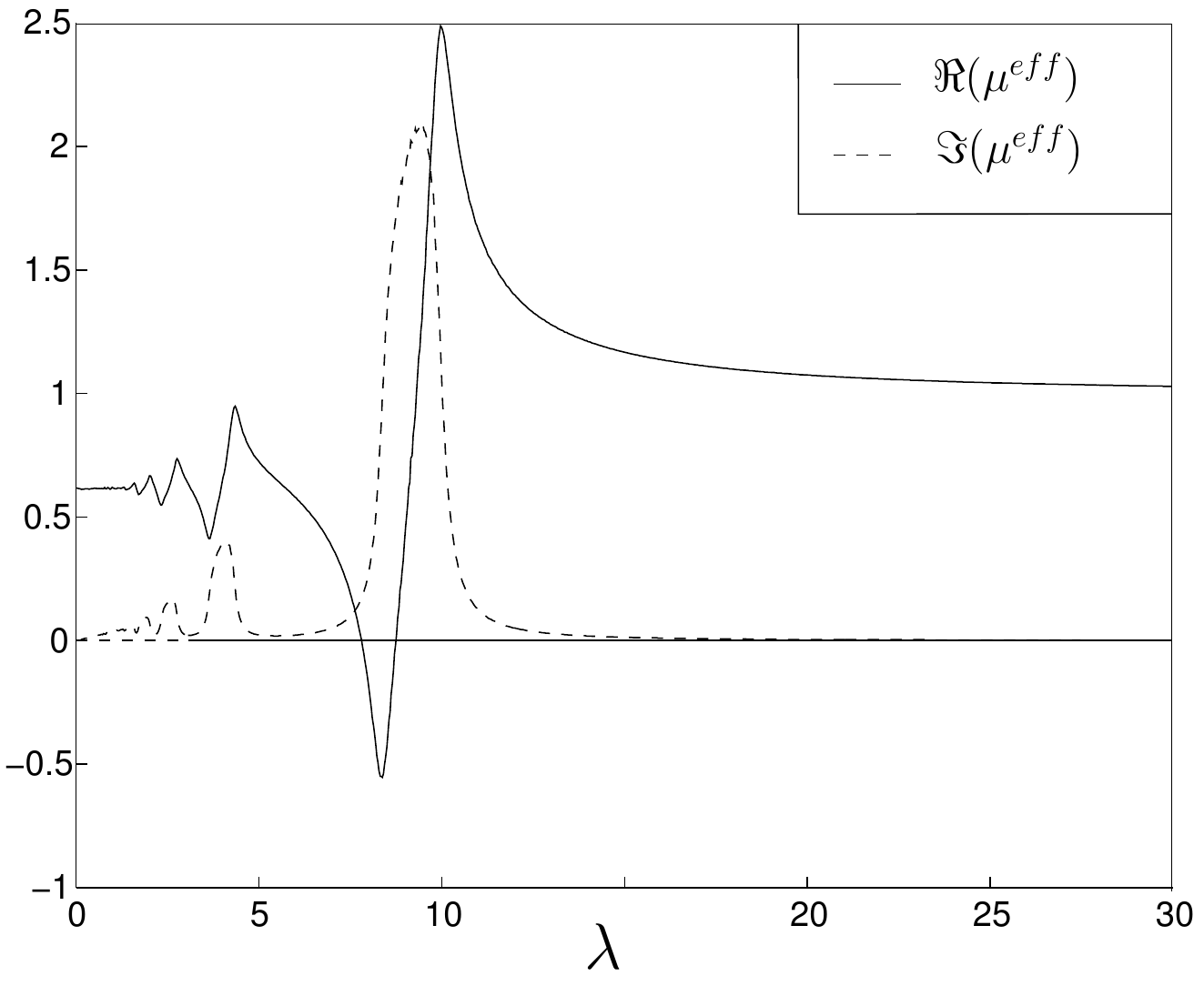}
\end{minipage}
\caption{\small Resonances in determinist (left) and random (right) cases. 
\label{fig.graph}}
\end{figure}

In Figure \ref{fig.graph}, we have drawn  the real and imaginary parts of $\muef$ predicted by the
homogenization formula  in the determinist case (left hand side)  and in the random case respectively (right hand side).
The abscissa  $\lambda=\frac{2\pi}{k_0}$ represents the normalized wavelength.

\subsection{The case of vanishingly small loss dielectric}
We  discuss now  about the dependence of the effective problem with respect to the permittivity law, in particular in the case where this law $\p$ is supported on the real axes (non dissipative medium) possibly violating the condition (\ref{eq.hyp}) needed for the validity of our homogenization result. For the sake of simplicity we take  $\p$ of the form
\begin{equation}\label{loi.initiale}
\p(\theta,\rho,\eps)=\delta_0(\theta-\theta_0)\otimes \gamma(\rho) \otimes g(a)\,da \otimes\delta_0(b)\ ,
\end{equation}
where $\eps=a+ib$, $g$ is a Lipschitz function on $\Rset$ with compact support and $\gamma$ 
is an arbitrary probability law for $\rho$ on $(0,\rho^+]$ (such that $B(\theta_0,\rho^+) \subset\subset Y$).
Then we introduce a small absorption parameter $h>0$ (designed to tend to zero) and set
\begin{equation}\label{e.def_ph}
\p_h:=\ \delta_0(\theta-\theta_0)\otimes \gamma(\rho) \otimes g(a)\,da \otimes\delta_0(b-h)\ .
\end{equation}
Clearly, for every $h>0$ the condition (\ref{eq.hyp}) holds and  $\p_h\stackrel{*}{\rightharpoonup} \p$ as $h$ tends to zero.
%
%
%
%
%
%
%
%
%
For every $h>0$, the condition (\ref{eq.hyp_im}) holds and by applying Theorem  \ref{main}, the homogenized medium obtained in the limit $\eta\to 0$ is
 characterized by the effective permeability \begin{equation}\label{e.intM_prin}
\muef_h(k_0):=1+\sum_n |c_n|^2\,  I_{h,n} \quad,\quad I_{h,n}:=\int_M\frac{\eps k_0^2\rho^4}{\lambda_n-\eps k_0^2\rho^2}\ p_h(d\theta\, d\rho\, d\eps).
\end{equation}

Our goal is to propose an effective theory for non dissipative random structures, like those described by (\ref{loi.initiale}),
by passing to the limit $h\to 0$. 
For every $n\in\Nset$, we introduce the real compactly supported Lipschitz function $f_n:\Rset\mapsto\Rset_+$ defined by
\begin{equation}\label{hn(s)}
f_n(s)\ :=\ \frac{\lambda_n}{k_0^2} \ \int_{(0,\rho^+]} g\Big(\frac{\lambda_n}{k_0^2\, \rho^2} + s\Big)\ \gamma(d\rho)\ .
\end{equation}

%

\begin{figure}[!h]
\begin{minipage}[b]{0.48\linewidth}
\includegraphics[width=\linewidth]{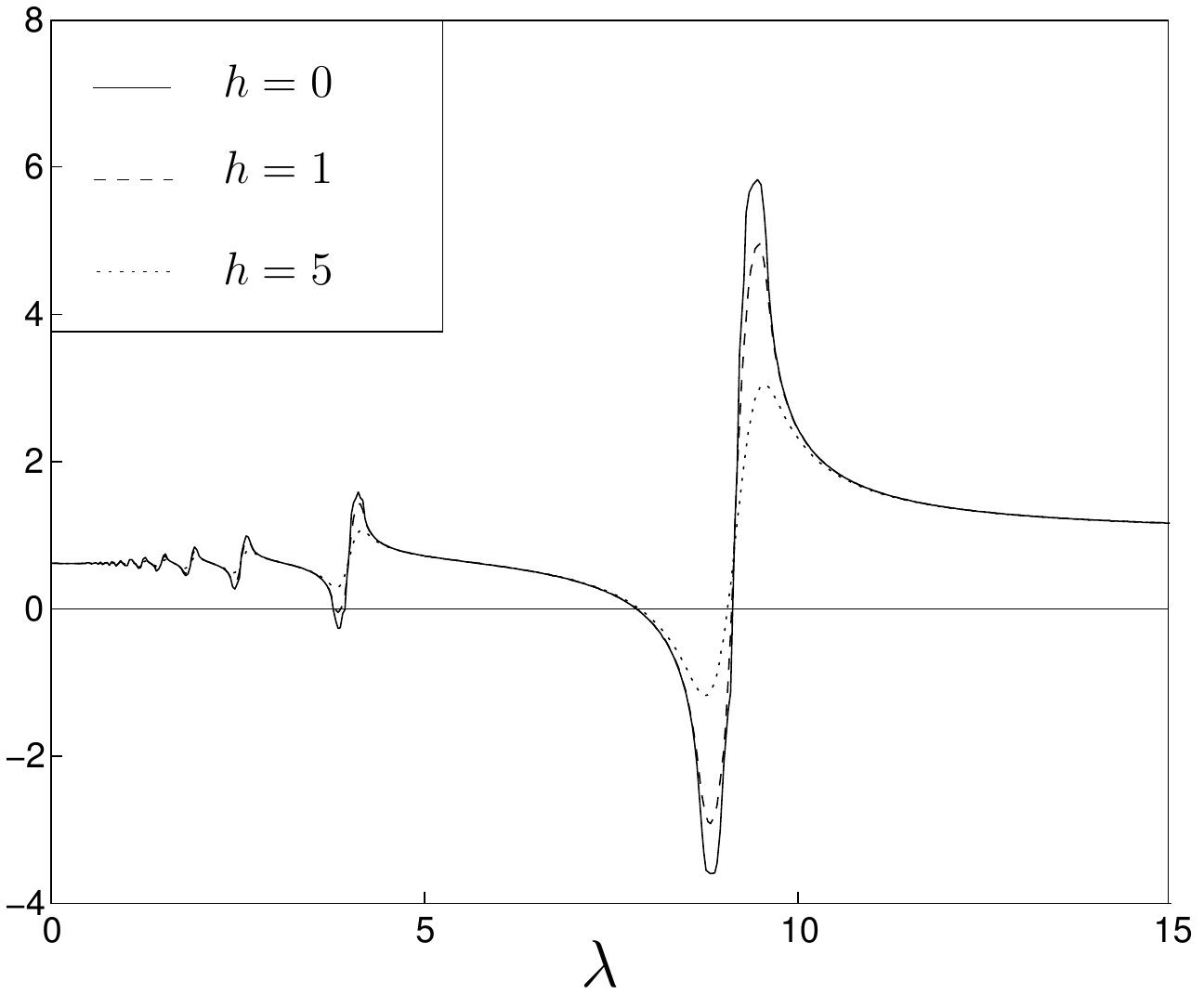}
\end{minipage}
\begin{minipage}[b]{0.48\linewidth}
\includegraphics[width=\linewidth]{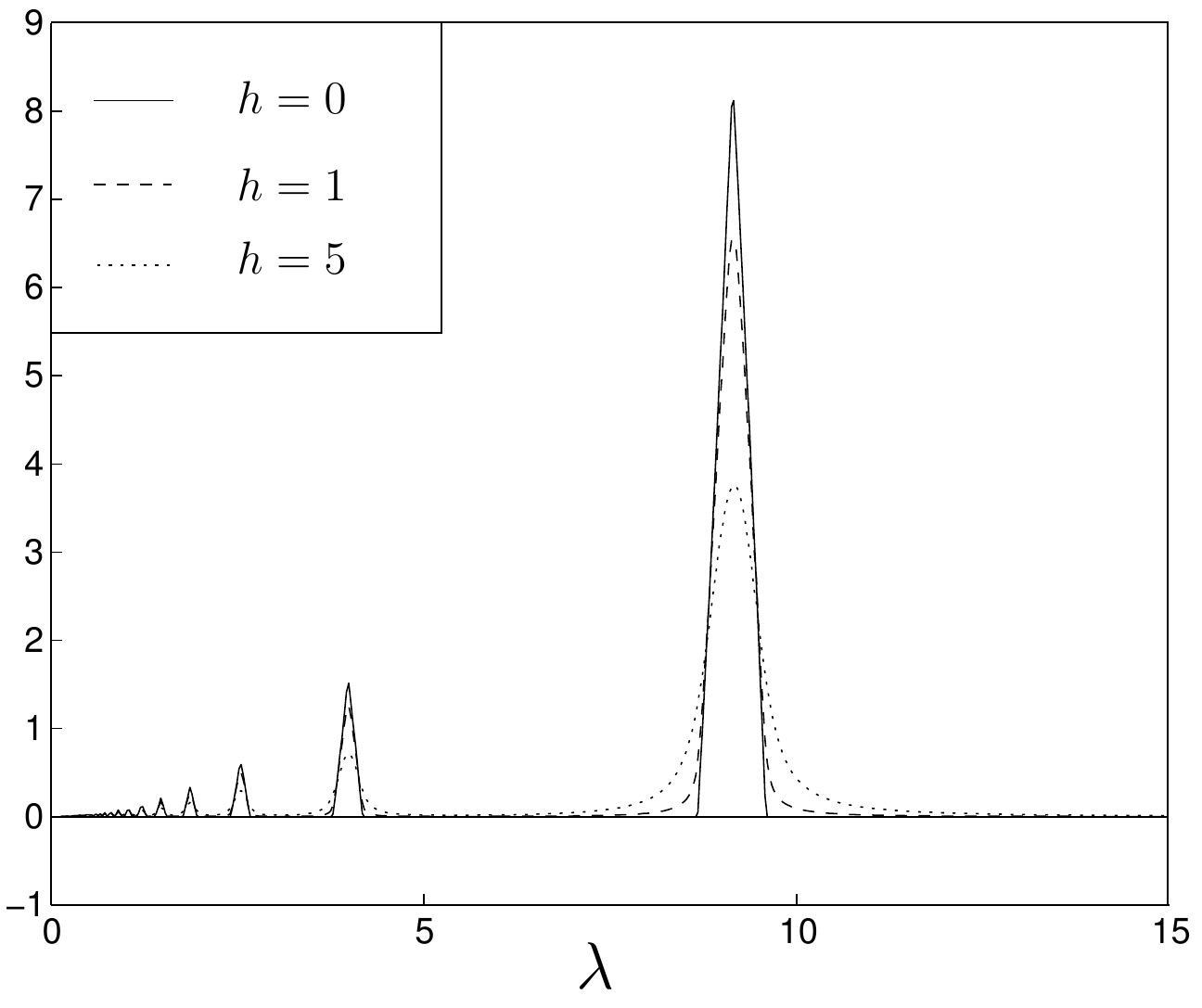}
\end{minipage}
\caption{\small Real and imaginary part of $\muef_h$ for $h\in \{0,1,5\}$
 \label{fig.dis_lim}}
\end{figure}

\begin{proposition}\label{th.lim_Ih}
For every $k_0 \in \Rset^+$, there holds $\muef_h(k_0)\to \muef(k_0)$ where
\begin{equation}\label{e.mu_lim}
\mu^{\mathrm{eff}}_0(k_0)=1 - \pi \int_{(0,\rho+]} \rho^2 \, \gamma(d\rho) + 
\sum_{n\in\Nset} |c_n|^2\, \left(-  \mathrm{PV} \int \frac{f_n(s)}{s}\, ds \, +\, i\, \pi \, f_n(0)\right),
\end{equation}
(where $\mathrm{PV}(\int f)=\dis\lim_{s\to 0} \textstyle\int_{\Rset\setminus [-s, s]} f(t)\,\mathrm{d} t$ and refers to the Cauchy principal value).

\end{proposition}
Some  comments  about  the validity of the resulting model can be found in Remark \ref{noncommute}. 
In order to illustrate the random dependance,
we draw in figure \ref{fig.dis_lim}  the function $\muef_h$ for decreasing values of parameter $h$ 
and  $\muef_0$ given in (\ref{e.mu_lim}) which corresponds to $h=0$.
To simplify, we choosed the law of $\gamma(\rho)$ to be Dirac mass at $\rho_0=0.35$ while
 $g(a)= 10^{-2} \big(10 - |a- 100|\big)^+$.


\begin{remark} \label{noncommute}\upshape{The limit absorption process described in Proposition \ref{th.lim_Ih} is stable in the sense that the result remains unchanged if
in (\ref{e.def_ph}) we substitute $\delta_0(b-h)$ with $\frac1{h} \zeta(\frac{b}{h})$,
being $\zeta$ any probability law  on $\Rset^+_*$ compatible with (\ref{eq.hyp}). 
At this point it is tempting to claim that that, even  if the initial random law in the rods is non dissipative,  the limit medium as $\eta\to 0$ can still be described by the homogenization theory with an effective  permeability law given by (\ref{e.mu_lim}).

However the situation is not so simple and we face here a paradox. Indeed,
 if the initial probability law is non dissipative like in  (\ref{loi.initiale}),  it looks obvious that 
 the homogenized limit should be  described by real effective coefficients.
Surprisingly it is not the case in our model: according to formula (\ref{e.mu_lim}) the imaginary part of $\muef_0$ shows positive 
 values when the incident wave number $k_0$ ranges in suitable intervals of $\Rset^+$.
 Several possible explanations can be addressed. First
  our asymptotic analysis $\eta\to 0$ has been performed  in harmonic regime fixing a priori the frequency of the incident wave. 
Second we applied  the limit absorption principle ($h\to 0$) 
once the microscale parameter $\eta$ was sent to zero. This clearly is  only one strategy
 among many others. 
}

\end{remark}

\begin{remark}\label{rem.conv_dissipa} \upshape{If $\Im(\muef_0(k_0))>0$, Proposition \ref{th.lim_Ih} can be improved as follows:
let  $u_h$ be the solution of (\ref{e.limiteq}) being in (\ref{globalepsmu}) $\muef$ substituted with $\muef_h$.
Then $u_h\to u $   strongly in $L^2_{\loc}(\Rset^2)$ where $u$ is the unique solution of (\ref{e.limiteq})
with  $\muef_0$ in (\ref{globalepsmu})  determined by (\ref{e.mu_lim})
(the main point of the proof is an $L^2$ uniform estimate for $\{u_h\}$ which   can be obtained by using the same kind of contradiction argument as in the proof of Theorem \ref{main}).
}
\end{remark}

\paragraph{Proof of proposition \ref{th.lim_Ih}}
Let us  plug in (\ref{e.intM_prin}) the expression (\ref{e.def_ph}) of $p_h$ . By  using the change of variable $s=a- \frac{\lambda_n}{k_0^2\rho^2}$ together with Fubini's formula, we are led to
\begin{equation}\label{Ihn}
I_{h,n}\ =\   \int_{(0,\rho+]} \rho^2 \, \gamma(d\rho) -     \int_{-\infty}^{+\infty}  \frac{ s}{s^2+h^2}\ f_n(s)\ ds \ +\
i\, \int_{-\infty}^{+\infty} \frac{ h}{s^2+h^2}\ f_n(s)\ ds\, .
\end{equation}
Here $f_n$ given by (\ref{hn(s)}) inherits the Lipchitz continuity of $g$, and therefore, for every $n$, there holds:
$$
\lim_{h\to 0} \int_{-\infty}^{+\infty}  \frac{ s}{s^2+h^2}\, f_n(s)\, ds \ =\  \mathrm{PV} \int \frac{f_n(s)}{s}\, ds \ , 
$$
$$
 \lim_{h\to 0} \int_{-\infty}^{+\infty}  \frac{ h}{s^2+h^2}\, f_n(s)\, ds \ =\ \pi \, f_n(0)\ .
$$
Thus we have
\begin{equation} \label{limIhn}\lim_{h\to 0} I_{h,n}= I_n := \int_{(0,\rho+]} \rho^2 \, \gamma(d\rho)  - \mathrm{PV} \int \frac{f_n(s)}{s}\, ds \, +\, i\, \pi \, f_n(0)\, .\end{equation} 
In view of (\ref{e.mu_lim})(\ref {limIhn}) and recalling that $\sum_n |c_n|^2= \pi$, the convergence (\ref{e.mu_lim}) is established provided we have
  $\dis \lim_{h\to 0} \big( \sum_n |c_n|^2 I_{h,n}\big) = \sum_n |c_n|^2 I_n.$
In order to prove previous equality as well as the summability of the series,
it is enough to show the following uniform upper bound for suitably large  $n_0$:
\begin{equation}\label{claim0}
 \sup \left\{ |I_{h,n}| \ :\ n\ge n_0\ ,\ h>0\right\} \ <\ +\infty
 \end{equation}
 Let us set \ $\alpha_n:= \inf \{ |s|\ :\ f_n(s)\not=0 \}$. 
 Looking at  (\ref{hn(s)}), since  $\rho $ ranges in $[0,1/2)$ and $g$ is compactly supported while $\lambda_n\to\infty$ as $n\to\infty$,
 it is easy to check that 
 $$\liminf_{n\to\infty} \ \frac{\alpha_n}{\lambda_n}\ \ge\ \frac4{k_0^2}\, .$$
 On the other hand, from the expression in (\ref{Ihn}), we deduce that
 $$ |I_{h,n}|\ \le\ \frac1{4} + \frac1{\sqrt{\alpha_n^2 + h^2}} \int f_n(s)ds\ \le\ \frac1{4} + \frac{\lambda_n}{k_0^2 \alpha_n}\ .$$
 The upper bound (\ref{claim0}) follows and the proof is finished.
 \QED

\section{Stochastic framework}
The random set $\DomD_\eta(\omega)$ defined in \eqref{def:Deta} can be generated (after a proper rescaling) through a continuous dynamical system.
To that aim we will follow basically the approach developed in
\cite{JKO} (for the homogenization of randomly perfored domains) and more recently in \cite{ZP}. 

\paragraph{Dynamical system and ergodicity}

On our probability space $(\spPt, \mathcal{A},\PPt) $, we define a group of transformations $T_x:\spPt\to \spPt,\ x\in \Rset^2$ as follows
\begin{equation}\label{mapT}
T_x\left((\mi_j)_{j\in\Zset^2},y\right)=\left((\mi_{j+[x+y]})_{j\in\Zset^2},x+y-[x+y]\right),
\end{equation}
where by abuse of notations we write $[x]$, for $x\in\Rset^2$, meaning the couple $([x_1], [x_2])$ formed by the integer parts of the coordinates of $x$.

\med
It is easy to check that the map $(x,\omega)\mapsto T_x(\omega)$ is  Borel regular
  and that it enjoys the usual properties of a {\em ergodic dynamical system}:
 for every $x\in\Rset^2$, $T_x$ preserves the measure $\PPt$. 
Moreover since the discrete dynamical system associated with the Bernoulli shift over the product space $(\spP, \PP):=(\spM^{\Zset^2}, \bigotimes_{\Zset^2}\p)$  
 is ergodic (see for instance \cite[Theorem 8.4.5]{Dud}), it can be easily checked (as noticed in \cite{ZP}) that $T_x$ is ergodic on $(\Omega,\mathcal{A},\PPt)$. This means that, for every $x\in \R^2$, one has $\PPt(A)\in\{0,1\}$
for every $T_x$-invariant  $A\in \mathcal{A}$ i.e. such that $\PPt(T_xA \vartriangle A)=0.$ 
 
\med As a fundamental consequence, by  Birkhoff ergodic theorem (see \cite[Proposition 12.2.II]{DVJII}), for every $f\in L^1(\Omega,\mathcal{A}, \mathbb{P})$, the following equality holds  almost
surely 
\begin{equation}\label{ergomean}
\lim_{r\to\infty}\frac{1}{\mathcal L^2(B_r)}\int_{B_r}f(T_x\omega)dx=\int_{\Omega} fd\mathbb{P},\qquad \mathbb{P}-\text{a.s.}\ .
\end{equation}

%
%
%
\med
%


\paragraph{Statistically homogeneous random fields and scaling}
With our dynamical system, we associate the family of \textit{statistically homogeneous random fields} which are functions $f(x,\omega)$ of the kind $F(T_x \omega)$, being $F$ a  measurable function on $\Omega$. The rescaled versions $F(T_{\frac{x}{\eta}} \omega)$
 can be seen as a stochastic counterpart of the class of periodic functions 
 oscillating at scale $\eta$.
Such random fields are well fitted to represent our random permittivity    
because the properties of a rescaled cell $Y^j_\eta(\omega)=\eta(j-y(\omega)+Y)$ in the configuration $\omega$ can be expressed in terms of the reference cell $\tor$ in the configuration $T_{\frac{x}{\eta}}\omega$.
 More precisely, recalling the definition (\ref{Sigma}) of $\Sigma$, the random sets $\DomD_\eta^j(\omega)$ defined in (\ref{def:Deta}) can be characterized as follows 
\begin{equation} \label{eq.xinDeta}
x\in \DomD_\eta^j(\omega)\ \iff\ j =\left[\frac{x}{\eta}+\boldsymbol{y}(\omega)\right]\ \text{ and } \ (x,T_{\frac{x}{\eta}}\omega)\in \mat\times\Sigma.
\end{equation}
Accordingly, since $\boldsymbol{\varepsilon}_j( \omega)=\boldsymbol{\varepsilon}_0(T_\frac{x}{\eta} \omega)$ on each $D^j_\eta(\omega)$,
the inverse of the random relative permittivity $a_\eta(x,\omega)$
reads:
\begin{equation} \label{eq.a_eta_fin}
a_\eta(x,\omega)=1_\mat (x)\left(\frac{\eta^2}{\boldsymbol{\varepsilon}_0(T_\frac{x}{\eta} \omega)}1_{\Sigma}\big(T_\frac{x}{\eta}\omega\big)+1_{\Sigma^*}\big(T_\frac{x}{\eta}\omega\big)\right)+ {1}_{\Rset^2\setminus\mat }(x)
\end{equation}

\paragraph{Stochastic derivative}

We associate with  $T_x$ a group of transformations $U_x$, $x\in\Rset^2$ acting on measurable functions on $\spPt$, i.e. if $f:\spPt\to \Rset$ is $\mathcal A$-measurable, then
\[
(U_xf)(\omega)=f(T_x \omega),\quad \omega\in \spPt,\quad x\in \Rset^2.
\]
\begin{lemma} \label{p.strcont} For every $x\in \R^2$,  $U_x$ is a unitary transform on $L^2(\spPt,\PPt)$
and the group $x\in\R^2 \mapsto U_x$ is strongly  
 continuous.
\end{lemma}
\proof
Clearly  $U_x$ is unitary in $L^2(\spPt,\PPt)$ since $T_x$ preserves the measure $\PPt$.
 Let $f\in L^2(\spPt,\PPt)$, $\varepsilon>0$ and let $g\in C(\spPt)$ such that $\|f-g\|_{L^2(\spPt,\PPt)}<\varepsilon$.
 Then,  for any $x\in \Rset^2$, we have
 \[
 \|U_xf-f\|_{L^2(\spPt,\PPt)}\leq 2\varepsilon+ \|U_xg-g\|_{L^2(\spPt,\PPt)}.
 \]
It is therefore sufficient to show the continuity result assuming that $f$ is bounded continuous in $\Omega$.
For $y\notin \partial\tor $ and for $x$ small enough such that $[x+y]=0$, it holds
\begin{eqnarray*}
U_x f(\omega)
=f\big((\mi_{z})_{z\in\Zset},x+y\big).
\end{eqnarray*}
Since $f$ is continuous, this implies that $U_x f(\omega)\to f(\omega)$ as $|x|\to 0$, whenever $\omega=(m,y)$ is
such that $ y\notin\partial\tor$, that is $\PPt$ a.e. $\omega\in \Omega$.
By dominated convergence we conclude that
\[ \lim_{x\to 0}\ 
\int_\spPt |U_x f(\omega)-f(\omega)|^2\, d\PPt
\ =\ 0\]
\QED

\med
The following notion of stochastic derivative is taken from \cite{ZP}.
\begin{definition}[Stochastic partial derivative]
The stochastic partial derivative $\partial^s_i$ (i=1,2) is the infinitesimal generator of the strongly continuous semigroup $U_{te_i},t\geq 0$ in the space $L^2(\Omega)$.
For $i=1,2$, the domain $D(\partial^s_i)\subset L^2(\Omega)$ of $\partial^s_i$ is given by
\begin{equation}\label{domaingrad}
D(\partial^s_i)=\Big\{f\in L^2(\spPt;\PPt)\ :
\ \DS \frac{U_{te_i}f-f}{t}\,  \text{ converges in  $L^2(\spPt;\PPt)$ as $t\to 0$} \Big\}
\end{equation}
and for $f\in D(\partial^s_i)$, we set \ $\DS \partial_i^s f(\omega)=\lim_{t\to 0} \frac{U_{te_i}f(\omega)-f(\omega)}{t},\ \omega\in \spPt.$

Accordingly the  stochastic gradient $\Grads:D(\Grads)\to (L^2(\Omega;\PPt))^2 $ is defined by
$$
\nabla^s f:=(\partial_1^sf,\partial_2^sf),\qquad f\in D(\Grads):= D(\partial_1^s)\cap D(\partial_2^s).
$$
\end{definition}
Notice that such a gradient has a vanishing average over $\Omega$. Moreover, for $f,g \in D(\Grads)$, one has $ \E(\partial_i^sf \, g) =- \E(f \, \partial_i^sg).$

\paragraph{Approximation} A function in $\LOP$ can be approximated by elements of $D(\Grads)$ by using the following analogue of the ``convolution'' procedure.
Let $\Psi_n(x) :=n^2\, \Psi(nx)$ where $\Psi:\R^2\to \R^+$ is a fixed smooth regularisation kernel with compact support such that $\int_{\R^2} \psi =1$. 
Then for every $f\in \LOP$, we set
\begin{equation}\label{convol}
f_n(\omega)  \ :=\ \int_{\R^2} f(T_x \omega) \, \Psi_n(x)\, dx 
\end{equation}
 \begin{lemma}\label{closable} \vskip0.5cm
For every $f\in\LOP$, $f_n $ given in \eqref{convol} belongs to $D(\Grads)$  and converges strongly to $f$ in $\LOP$.
Moreover it holds
\begin{equation}\label{convolution}
 \nabla^s \, f_n (\omega) = -\int_{\R^2} f(T_x\omega) \, \nabla \Psi_n(x) \, dx \quad,\quad \PPt\ {\rm a.e.}\ \omega\in \Omega 
\end{equation}
As a consequence the space $D(\Grads)$ is dense in $\LOP$ and the operator $\Grads$ is closable. 
\end{lemma}
\proof By the change of variable $x\mapsto x+t\, e_i$, we have for $t>0$ and   $i\in\{1,2\}$
$$  \frac{f_n (T_{ t \, e_i} \omega)- f_n(\omega)}{t}  = 
\int_{\R^2} f(T_x\omega) \,   \frac{(\Psi(x-t\, e_i)- \Psi_n(x))}{t} \, dx \ .$$ 
For small $t$, the integral above can be restricted to a compact neighborhood  $K$ of the support of $\Psi$. Then,
since $\int_{K\times\Omega} |f(T_x \omega)|^2 \, dx\, d\PPt = |K|\, E(|f|^2)<+\infty$, by  using Cauchy-Schwartz inequality, we easily derive that 
$$\frac{f_n (T_{ t \, e_i} \omega)- f_n(\omega)}{t} +\int_{\R^2} f(T_x\omega) \, \frac{\partial\Psi_n}{\partial x_i} \, dx \ \to \ 0 \qquad
\text{ strongly in $\LOP$} \ .$$
Thus $f_n\in D(\Grads)$ and \eqref{convolution} holds. 
On the other hand it is easy to check that $f_n\to f $ in $\LOP$. 
Therefore  we have shown that the operator $\Grads$ is densily defined. 
 To check that it is closable  we consider $u_n$ in $D(\Grads)$ such that $u_n\to 0$ and $\nabla^s u_n \to \xi$ in $\LOP$. For every  $f\in D(\Grads)$, one has:
$ \E( \xi_i\, f)  \, =\, \lim_{n}\,  \E( \partial_i^s u_n\, f)  =- \lim_{n}\,  \E( u_n \, \partial_i^s\, f) = 0  ,$\
thus $\xi_i=0$ by the density of $D(\Grads)$. 
\QED

\paragraph{Stochastic Sobolev space}
The Sobolev space $H^1_s(\spPt)$ will be the completion of $D(\Grads)$ with respect to the Hilbert norm
$
\|f\|_{H^1_s}= \|f\|_{L^2(\spPt;\PPt)}+\|\nabla^sf\|_{L^2(\spPt;\PPt)}.
$
By Lemma \ref{closable}, $D(\Grads)$ is closable and admits a unique extension to $H^1_s(\Omega)$. This extension still denoted $\Grads$
acts as a linear continuous operator from $H^1_s(\Omega)$ into $\LOP$.
We may now define a divergence operator by duality. Let
\begin{equation*}
 D(\Divs):= \Big\{f\in (\LOP)^2\ :\ \exists C>0\ , \E(f\cdot \nabla^s g)\le \ C \|g\|_{\LOP}\quad \forall g\in H^1_s(\Omega)\Big\}\ .
\end{equation*}
Then, for every $\sigma\in D(\Divs)$, the stochastic divergence $\Divs \sigma$ is characterized by the indentity
\begin{equation} \label{eq.ibp}
 \E( (\Divs \sigma) \,  g) \ =-\E(\sigma\cdot \nabla^s g) \quad  \text{for all $g\in H^1_s(\Omega)$}.
 \end{equation}
Equipped with the norm $ \sigma \to [\E \left(|\sigma|^2 + |\Divs \sigma|^2\right)]^{1/2}$, $D(\Divs)$ is a Hilbert space as well.
The subspace $L^2_{\rm sol}(\Omega)$ of ``solenoidal'' fields will play an important role. It is defined as
\begin{equation}\label{solenoid}
L^2_{\rm sol}(\Omega) \ =\ \{ \sigma \in D(\Divs)\ :\ \Divs \sigma = 0 \}. 
\end{equation}

We are now in position to give a characterization of Sobolev space $H^1_s(\Omega)$ and $D(\Divs)$.
\begin{proposition}\label{Sobequiv} The following assertions are equivalent:
\begin{enumerate}
\item  [i)]\ $f\in H^1_s(\Omega)$
\item [ii)]\ There exists a sequence $f_n\in H^1_s(\Omega)$ such that \\
$$f_n\to f \quad \text{in $\LOP$\quad and }
\quad \sup_n \ \E(|\nabla^s f_n|^2) <+\infty\ .$$
\item [iii)]\ There exists a constant $C>0$ such that \ $ \E( f\, \Divs \sigma) \le C\, \| \sigma\|_{(\LOP)^2}$, \ for every $\sigma\in D(\Divs).$
\item [iv)]\ For a.a. $\omega$,  the realization  $x\mapsto f(T_x\omega)$ belongs to $W^{1,2}_{\rm loc}(\Rset^2)$  
and its distributional gradient  $\nabla_x (f(T_x\omega))$ is a statistically homogeneous random field in $(L^2_{\rm loc}(\R^2\times\Omega))^2$.

\end{enumerate}
In all these cases, for a.a. $\omega$, we have the following relation holding in the distributional sense in $\R^2$ :
\begin{equation} \label{eq.chainrule}
\frac{\partial }{\partial x_i}\big(f(T_x \omega)\big)=\left(\partial_i^sf\right)(T_x \omega).
\end{equation}

\end{proposition}

\proof The equivalence between assertions i), ii) and iii) is straightforward once 
it is observed that
the graph $G :=\{ (f, \nabla^s f) \ :\ f\in H^1_s(\Omega)\}$ is a closed subspace of $\LOP \times (\LOP)^2$
whose orthogonal reads\ $G^\perp :=\{ (\Divs \sigma, \sigma)\ :\ \sigma\in D(\Divs)\}.$ \\
Let us  prove now that every element $f\in H^1_s(\Omega)$ satisfies the property iv) together with \eqref{eq.chainrule}. 
In a first step, we assume that $f\in  D(\Grads)$. Then for every pair  $(\varphi, g)\in C^\infty_c(\R^2)\times \LOP$ and $i\in\{1,2\}$,
by  the invariance property of $\PPt$, one has 
\begin{eqnarray}\nonumber
\E\left(g(\omega) \int \partial_i^s f(T_x \omega) \,\varphi(x) \, dx    \right) &=& \lim_{t\to _0} 
\int_{\Omega}\int_{\R^2}   \frac{ f(T_{x+t e_i} \omega) -  f(T_{x} \omega)}{t} \varphi(x)\,g(\omega) \, dx\, d\PPt \\
\nonumber
&=& \lim_{t\to _0} 
\int_{\Omega}\int_{\R^2}  f(T_{x} \omega) \frac{ \varphi (x)-\varphi(x-t e_i)}{t}  g(\omega) \, dx\, d\PPt \\ \label{equivH1}
&=&  - \int_{\Omega}\int_{\R^2}  f(T_{x} \omega) \partial_{x_i} \varphi(x)\,  g(\omega) \, dx\, d\PPt \,
\end{eqnarray}
from which follows that, a.a. $\omega$,  the map  $x\mapsto f(T_x\omega)$ belongs to $W^{1,2}_{\rm loc}(\Rset^2)$  with \eqref{eq.chainrule}.

%
%
%

In a second step, we consider an approximation sequence $f_n\to f$ in $H^1_s(\Omega)$ such that $f_n \in D(\Grads)$. By the first step, may apply \eqref{equivH1} to $f_n$ which, after the change of variable $\omega\mapsto T_{- x}\,\omega$, leads to
\begin{equation}\nonumber
\E\left(\partial_i^s f_n( \omega)  \int_{\R^2} \varphi(x) \, g(T_{-x}\omega) \, dx  \right) =  - \E\left(  f_n( \omega)\int_{\R^2} \partial_{x_i} \varphi(x)\,  g(T_{-x}\omega) \, dx\, \right)\ .
\end{equation}
The relation above remains true for $f$ by passing to the limit $n\to\infty$. Then by applying  the inverse change of variable $\omega\mapsto T_{- x}\,\omega$
and by the arbitrariness of $g\in\LOP$, we conclude that \eqref{eq.chainrule} holds true for every $f\in H^1_s(\Omega)$.

Conversely assume that iv) holds and let us prove that $f\in H^1_s(\Omega)$. 
 We consider the approximation $f_n$ defined in \eqref{convol}. By Lemma \ref{closable}, we have $f_n\in D(\Grads)$ and, since by hypothesis $f(T_x \omega)$ 
 belongs to $W^{1,2}_{\rm loc}(\Rset^2)$ we may rewrite \eqref{convolution} as
 $ \Grads f_n(\omega) = \int  \nabla_x [f(T_x\omega)] \, \Psi_n(x)\, dx.\ $
 Then by Jensen inequality (recall  $\int\Psi_n=1$), we derive that
 $$ |\Grads f_n(\omega) |^2 \ \le\ \int  \Big |\nabla_x [f(T_x\omega)] \Big|^2 \, \Psi_n(x) \, dx$$
 Now if $\nabla_x [f(T_x\omega)]$ is statistically homogeneous i.e. of the kind $F(T_x \omega)$ for a suitable $F\in (\LOP)^2$,
  we infer that $\sup_n \E(|\Grads f_n(\omega) |^2) \le \E(|F|^2)  < +\infty$.
 The conclusion follows by applying assertion ii) to the sequence $\{f_n\}$. 
\QED

\begin{corollary}\label{stochdiv} Let $\sigma \in (\LOP)^2$ and $h \in \LOP$. Then $\sigma$ belongs to $D(\Divs)$ with $ \Divs \sigma=h$  if and only if, for
a.e. $\omega\in\Omega$, the function $x\mapsto \sigma(T_x \omega)$ satisfies ${\rm div}_x (\sigma(T_x \omega)) =h(T_x \omega)$ in the distributional sense in $\R^2$. \end{corollary}
\proof We set $\sigma_\f(\omega):= \int_{\R^2} \f(x)\, \sigma(T_x\omega) \, dx.$ for given $\sigma \in (\LOP)^2$ and $\varphi\in C^\infty_c(\R^2)$. 
Then in a similar way as in \eqref{convolution},  we have that $\sigma_\varphi\in D(\Divs)$ with
\begin{equation}\label{stepuno}
\text{ for\,  $\PPt$  a.e. $\omega\in \Omega$} \quad ,\quad   \Divs \sigma_\f(\omega) \ =\ -\int  \sigma(T_x\omega)\cdot \nabla\f(x) \, dx
\end{equation}
Indeed, for  every $g\in H^1_s(\Omega)$, one has $ \nabla_s g(T_{-x} \omega)=- \nabla_x  (g(T_{-x}\omega))$ and by using Fubini and  changes of variable $\omega\mapsto T_{\pm x}\,\omega$, one gets
\begin{align*}
 \E( \sigma_\f \cdot \nabla^s g)&=  \E \left(\int \sigma(T_x \omega)\cdot  \nabla^s g(\omega) \, \f(x) \, dx  \right)  \\
 & \hskip-1cm =
-  \E \left( \int  \sigma(\omega)\cdot \nabla_x (g(T_{-x}\omega)) \,  \f(x)\, dx\right) 
 =  \E \left( \int \sigma(\omega)\cdot \nabla \f(x)\, g(T_{-x}\omega) \, dx\right) \\
 &=  \E \left( g(\omega) \int \sigma(T_x \omega)\cdot \nabla \f(x)\, dx  \right)\ ,
\end{align*}
which according to \eqref{eq.ibp} is equivalent to \eqref{stepuno}.

\med  Let us assume firstly that for a.a $\omega$  ${\rm div}_x (\sigma(T_x \omega)) =h(T_x \omega)$ holds in the distributional sense.
 Then setting $\sigma_n :=\sigma_{\Psi_n}$ where $\Psi_n$ is the regularization kernel used in Lemma \ref{closable}, we infer from \eqref{stepuno}
 that $\Divs \sigma_n (\omega)  =  \int  h(T_x\omega)\,\Psi_n(x) \, dx.$ Thus  $ (\sigma_n, \Divs \sigma_n) \to (\sigma,h)$ in $L^2(\Omega)$.
 As by construction the operator $\Divs$ has closed graph, we conclude that $\Divs \sigma=h$.
 
\med  Assume conversely that $\Divs \sigma=h$ and let $\f\in C^\infty_c(\R^2)$. Then we claim that
\begin{equation}\label{stepdue}
\text{ for\,  $\PPt$  a.e. $\omega\in \Omega$}\quad,\quad  \Divs \sigma_\f(\omega) \ =\ \int  h(T_x\omega)\, \f(x) \, dx\, .
\end{equation} 
Combined with \eqref{stepuno} and since $\f$ is arbitrary, this implies that  ${\rm div}_x (\sigma(T_x \omega))$ and $h(T_x \omega)$ agree as distributions.
In view of \eqref{eq.ibp}, we consider an arbitrary $g\in H^1_s(\Omega)$. Noticing that, for all $x$, $\Divs(\sigma(T_x \omega)) = \Divs\sigma(T_x \omega)= h(T_x\omega)$ \, (for a.e. $\omega$), we have by using Fubini
$$
\E( \sigma_\f \cdot \nabla^s g) =  \E \left(\int \sigma(T_x \omega)\cdot  \nabla^s g(\omega) \, \f(x) \, dx\!  \right)  =
-   \E \left(g(\omega) \int h(T_x \omega)\cdot\, \f(x)\, dx\!\right) \,.
$$
The claim \eqref{stepdue} follows from the characterization  \eqref{eq.ibp}.
\QED

\paragraph{Back to our particular case}
Let us particularize previous definitions  in the case of the dynamical system introduced in (\ref{mapT}), 
where the events $\omega\in \Omega$ are of the form $\omega=(m,y)$ with $(m,y)= ((m_k)_{k\in\Zset^2},y)\in \spP\times Y$. 
It is convenient to associated with every $f= f(m,y)$ in  $\LOP$, the following continuation on $\Pi \times \R^2$:
\begin{equation}\label{periodize}
{\tilde f} (m,x) \ :=\ f\left((m_{k+[x]}), x-[x]\right) \ =\ f(T_x(m,0)) \ .
\end{equation}
We observe that, in the case of a function $f$ independent of $m$,  ${\tilde f}$ coincides with the $Y$-periodization of $f$.
On the other hand, the realization of a general random field $f$ can be recast from the identity
$ f(T_x(m,y)) = {\tilde f}(m, x+y)$.

\begin{proposition}\label{p.diff_stoc}\ 
\begin{itemize}
\item[i)] It holds  $f\in H^1_s(\spPt)$ if and only if $f$ belongs to $ L^2(\spP,\PP;W^{1,2}(\tor))$ and  
for $\PP$ almost all $m\in\spP$,  the function $\tilde f (m,\cdot)$ is an element of $W^{1,2}_{\rm loc}(\R^2)$.
In this case, one has
\begin{equation}\label{e.derivative}
\partial_i^sf(m,\cdot) =\frac{\partial f}{\partial y_i}(m,\cdot)\qquad \mbox{a.e in $Y$}.
\end{equation}
 
\item[ii)] A vector field $\sigma\in (\LOP)^2$ belongs to $L^2_{\rm sol}(\Omega)$ if and only if
its extension $\tilde \sigma(m,\cdot)$ is divergence free in $\R^2$ for almost all $m\in\spP$.

 \item[iii)] Let $f \in H^1_s(\spPt)$ be such that $\nabla^s f =0$  for $\PPt$-almost every $\omega\in\Sigma^*$.
 Then  $ f$ is constant $\PPt$-a.e. in $\Sigma^*$, that is there exists a contant $c$ ({\em independent} of $m$)  such that
 for  $\PP$ a.a. $m\in\spP$ 
\begin{equation}\label{constancy}
f(m, \cdot) = c \quad \text{ a.e. on  $\{y\in Y : |y-\btheta_0(m) | \ge \brho_0(m)\}$}\ . 
\end{equation}
   
\end{itemize} 


\end{proposition}

\begin{remark}\label{periodic}\rm

  According to the assertion i), a function $f(m,y)\in L^2(\Omega)$ such that $f=f(y)$ is independent of $m$
belongs to $H^1_s(\Omega)$ if and only if $f(y)\in W^{1,2}_\sharp(Y)$.
In the general case, checking that an element $f\in  L^2(\spP,\PP;W^{1,2}(\tor))$ satisfies $\tilde f (m,\cdot)\in W^{1,2}_{\rm loc}(\R^2)$
amounts to check a compatibility condition for the traces of $f(m,\cdot)$ on $\partial Y$, namely: 
\begin{equation}\label{compatible}
 f\big(T_{e_i}(m,t\,e_j)\big)\ =\ f(m,t\, e_j+e_i)\quad  ,\quad \text{a. e. $t\in(0,1)$ \ , \ $\{i,j\}=\{1,2\}$} .
\end{equation}
This holds true for instance if $f(m,\cdot)$ is in $W^{1,2}_\sharp(Y)$ with a trace 
 \textit{independent of~$m$}.
\end{remark}

\begin{remark}\rm
 The constancy property in $iii)$ will be crucial in the homogenization process. It will ensure that the internal resonances associated to each rod can be handled separately.
An abstract proof can be deduced directly from  the ergodicity of the dynamical system (see  \cite{ZP}). Here
in order to help for a more intuitive  understanding, we  proceed directly in our particular case.
\end{remark}

%
%
%
%
%
%
\proof Let $f\in H^1_s(\Omega)$. In view of the assertion iv) of Proposition \ref{Sobequiv}, we know that for $\PPt$ almost all $(m,y)$,
the function $x\mapsto \tilde f(m, x+y) (= f(T_x(m,y)) $ belongs to $W^{1,2}_{\rm loc}(\R^2)$. Since this property is invariant by translation,
it follows that for $\Pi$ a.e. $m$, it holds $\tilde f(m,\cdot)
\in W^{1,2}_{\rm loc}(\R^2)$. Furthermore, by \eqref{eq.chainrule}, we know that
$$\nabla_x \tilde f(m, x+y)= \nabla^s f(T_x(m,y))  \quad \text{ for a.a. $(x,y)\in \R^2\times Y$}$$
In particular, let us choose a Lebesgue point $x\in Y$ for both $\nabla_x \tilde f(m,\cdot)$ and $\nabla^s f(m,\cdot)$ 
and average the equality above on a small ball $\{|y|\le r\}$ with $r< dist(x,\partial Y)$ so that $T_x(m,y) = (m,x+y)$. Then by taking the limit  as $r\to 0$
we obtain \eqref{e.derivative} holding for a.e. $x\in Y$. As a consequence, $f\in L^2(\spP,\PP;W^{1,2}(\tor))$ with
\begin{multline*}
 \int_\spP \Vert f(m,\cdot)\Vert^2_{H^1(Y)}\, d\Pi \ =\  \int_\spP \left(\int_Y |f(m,y)|^2 + |\nabla^sf(m,y)|^2 \, dy\right)  d\Pi 
 =\Vert f\Vert^2_{H^1_s(\Omega)} <+\infty\ .
\end{multline*}
Conversely, if  for a.a. $m\in\Pi$, the function $\tilde f (m,\cdot)$ given in \eqref{periodize} belongs to $W^{1,2}_{\rm loc}(\R^2)$, then it is also the case for
the  translated maps $x\to \tilde f (m,x +y) = f(T_x(m,y))$ for every $y\in Y$. Moreover we have 
$ \nabla_x (f(T_x(m,y))=  F (T_x(m,y)) $ where $F= \nabla_x \tilde f$ is an element of
$(\LOP)^2$. Indeed as $\tilde f(m,\cdot)$ agrees with $f(m,\cdot)$ on $Y$, one has 
$$\E(|F|^2)\ =\ \int_{\Pi} (\int_Y |\nabla_x f|^2 dx) \,d\pi \ \le\ \int_\Pi \Vert f(m,\cdot)\Vert^2_{W^{1,2}(Y)} <+\infty\ .$$
Therefore the random field\ $(\omega,x)\mapsto \nabla_x f (T_x(\omega))$ is statistically homogeneous on $\Omega\times\R^2$ and
we  conclude that $f\in H^1_s(\Omega)$ by applying the assertion iv) or Proposition \ref{Sobequiv}.

\med
The assertion ii) is direct consequence of Corollary \ref{stochdiv}.
Let us prove iii). From the assertion i), we deduce that the function $f\in H^1_s(\Omega)$ is such that
for $\PP$ almost all $m\in \spP$, $ f(m,\cdot)$ as an element of $W^{1,2}(Y)$ has a vanishing gradient 
on the connected set $Y\setminus B\big(\btheta_0(m),\brho_0(m)\big)$.  Therefore there exits a contant $c(m)$
such that $f(m,\cdot)= c(m)$ a.e on $Y\setminus B\big(\btheta_0(m),\brho_0(m)\big)$. 
%
 In addition, in order to fit with the trace compatibility condition \eqref{compatible}, 
we need that for a.e $t\in(0,1)$ and for $\{i,j\}=\{1,2\}$  
$$
c(m)=f(m,e_i+t\,e_j) =\ f(T_{e_i}(m,t\,e_j))\ =\ f\big((m_{k+e_i})_k,t\,e_j\big)\ =\ c\big((m_{k+e_i})_k\big)
$$
From this  follows that the function $m\mapsto c(m)$ is shift invariant over $(M,\pi)$, thus  constant by ergodicity. This yields \eqref{constancy}.

\endproof

\section{Two-scale analysis of the diffraction problem}\label{sec.twoscale}

In this section and all along the rest of the paper, we will consider a subset $\tilde \Omega\subset \Omega$ of
 full measure (that is $\PPt(\Tilde \Omega)=1$), the elements of which 
will be refered as  ``typical events''. \\
For such $\tilde\omega\in \tilde\Omega$,
we ask first that:

\begin{itemize}
\item  property \eqref{ergomean} holds for all $f\in C(\Omega)$
(according to the terminology in \cite{ZP}, $x\mapsto T_x(\tilde \omega)$ is a ``typical'' trajector of the dynamical system)
 \item  for generic elements $f$ in a dense subset of $H^1_s(\Omega)$ the realizations $x\mapsto  f(T_x(\tilde\omega))$ belong to $W^{1,2}_{\rm loc}(\R^2)$ and satisfy the chain rule \eqref{eq.chainrule}
\item for $\Psi$ in a dense subset of $D(\Divs)$, it holds  $ \Divx\big(\psi(T_x(\tilde \omega)\big)= \Divs\psi(T_x\tilde \omega)$ in $L^2_{\loc}(\R^2)$.
\end{itemize}

We notice that the existence of such a set $\tilde \Omega$ of full measure is straightforward 
by exploiting the separability of $C(\Omega)$ and, in order to fit to two last items, by applying Proposition \ref{Sobequiv} (resp. Corollary \ref{stochdiv})
to a countable dense subset of elements $f$ in  the Hilbert space $H^1_s$ (resp. $\psi \in D(\Divs)$).

\subsection{Two-scale convergence} \label{2scale}

As our aim is  to obtain convergence results almost surely, we will use a recent variant of 
the notion of ``stochastic two-scale convergence in the mean''  introduced for the first time in 1994 in \cite{BMW}. 
Precisely we will use the notion of
``realizationwise two-scale convergence'' as introduced in \cite{ZP} in the context of the homogenization of random thin structures and singular
measures. The definition below is given
using a reference open ball $B_R$ of $\Rset^2$ (where in practice the radius $R$ is so large that $B_R\supset \mat$).

\begin{definition}[Stochastic two-scale convergence] \label{d.twoscale}
We say that a sequence $f_\eta\in L^2(B_R)$
two-scale converges to $f_0\in L^2(B_R\times\Omega,\mathcal L^2\otimes \PPt)$, and we write $f_\eta(x)\twoscale f_0(x,\omega)$ if for all 
$  \tilde\omega\in \tilde\Omega$ it holds
\begin{equation}\label{e.stsc}
\lim_{\eta\to 0}\int_{B_R} f_\eta(x)\, \varphi(x)\, \psi(T_{\frac{x}{\eta}}\tilde\omega)\, dx = \E \left( \int_{B_R} f_0(x,\cdot)\, 
\varphi(x)\, \psi(\cdot) \, dx \right)
\end{equation}
for any $\varphi\in C_c^\infty(B_R)$ and $\psi\in L^2(\Omega)$.
The sequence will be said to be  {\em strongly} two-scale convergent (denoted $f_\eta \stwoscale f_0$) if in addition
\begin{equation}\label{two-strong}
\limsup_{\eta\to0}\int_{B_R} |f_\eta(x)|^2 dx \ \le \ \int_{B_R\times\Omega} |f_0(x,\omega)|^2 \, dx\, \PPt(d\omega)
\end{equation}
\end{definition}
A justification of  definition \eqref{e.stsc}  is that any sequence $f_\eta$ has a two-scale converging subsequence provided it is bounded in 
$L^2(B_R)$. Notice that the existence of a two-scale limit implies that the left hand member of \eqref{e.stsc} is \textit{independent of the choice of $\tilde\omega\in\O$}.
In that case, in order to identify such a two-scale limit, it is enough to restrict to a particular event $\o\in\O$.

\med
The main feature of strong two-scale convergence is that we have the following product rule: 
\begin{equation}\label{product}
f_\eta \stwoscale f_0 \ ,\ g_\eta \twoscale g_0 \quad \Longrightarrow\quad f_\eta \, g_\eta \twoscale f_0\, g_0 \ .
\end{equation}
This applies in particular for $f_\eta$ of the kind $f_\eta(x)= \varphi(x)\, \psi(T_{\frac{x}{\eta}}\o)$
with $\o$  arbitrary \mbox{in $\O$}, $\varphi\in C(B_R)$ and $\psi\in L^2(\Omega)$ (in that case, there holds $f_\eta \stwoscale \varphi(x)\, \psi(\omega)$).
Furthermore the strong convergence $f_\eta \stwoscale f_0$ implies that
\[
\lim_{\eta\to0^+}\int_{B_R}\left| f_\eta(x) - f_0(x,T_{\frac{x}{\eta}}\tilde\omega)\right|^2dx =0.
\]
whenever $f_0$ is suitably regular.

\begin{remark} \label{twoscale.rem0}\upshape
If in  definition (\ref{mapT}) we choose $\p$ to be a Dirac mass (determinist case),
we fall on the  classical notion  of two-scale (resp. strong two-scale)  convergence  
which has been introduced  years ago in  \cite{Allaire,Nguetseng}.
\end{remark}

%

\begin{remark} \label{twoscale.rem2}\upshape
Let us emphasize  that the notion of stochastic two-scale convergence 
introduced above will be applied to \textit{to each particular realization
$f_\eta(x, \tilde \omega)$} of a sequence of random functions we wish to study,  providing then a two-scale limit $f_0(x,\omega, \tilde \omega)$ \textit{depending a priori of the realization $\tilde \omega$ and of the
associated extracted subsequence}.
This apparent drawback disappears in the context of effective media theory
when  the homogenization procedure provides a \textit{unique} two-scale limit
 $f_0(x,\omega)$.  In addition the property of strong two-scale convergence 
 introduced above involves only the spatial variable $x\in B_R$ and 
 in practice much easier to reach  than globally in the product space $B_R \times \Omega$. 

\end{remark}

\subsection{Differential characterization of the two-scale limits}

In this subsection, we fix an event $\tilde\omega\in \tilde\Omega$ and to simplify notions we denote:
$$ u_\eta(x)= u_\eta(x,\tilde\omega) \quad,\quad a_\eta(x)= a_\eta(x,\tilde\omega) \quad,\quad D_\eta= D_\eta(\tilde\omega)\ .$$
We assume a priori that the sequences $(u_\eta)$, $(\eta \, \nabla u_\eta)$ and $(1_{B_R\setminus\DomD_\eta}\, \nabla u_\eta)$
are bounded in $L^2(B_R)$ and satisfy
\begin{equation}\label{eq.twosc}
 u_\eta\twoscale u_0(x,\omega)\ , \qquad
 \eta\nabla u_\eta\twoscale P_0(x,\omega)\ ,\qquad
1_{B_R\setminus\DomD_\eta}\nabla u_\eta\twoscale \chi_0(x,\omega)\ ,
\end{equation}
for suitable functions $u_0\in L^2(B_R\times\Omega,\mathcal L^2\otimes\PPt)$ and  $\chi_0, P_0\in \big(L^2(B_R\times\Omega,\mathcal L^2\otimes\PPt)\big)^2$.

\begin{proposition} Let $u_0$, $P_0$ and $\chi_0$ be the two-scale limit defined in \eqref{eq.twosc}.
Then for $dx\otimes\PPt$-almost every $(x,\omega)\in{B_R}\times \Omega$ we have the following relations:
%
\begin{align}
P_0(x,\omega)=0&\qquad \mbox{in \ } ({\mat}\times\Sigma^*) \ \cup \left[(B_R\setminus\mat)\times\Omega\right]\ ,\label{eq.D0} \\
\nabla^s u_0(x,\omega)=P_0(x,\omega)&\qquad\mbox{in }\, {B_R}\times \Omega \label{eq.nablau0}\ ,\\
\chi_0(x,\omega)=0 & \qquad\mbox{in } {\mat}\times\Sigma\label{eq.chi02}\ ,\\
\Divs \chi_0(x,\omega)=0&\qquad \mbox{in }{B_R}\times \Omega \label{eq.chi0} \ .
\end{align}
\end{proposition}
\proof
For $\varphi\in C_c^\infty({B_R})$ and $\psi$ in a dense subset of $D(\Divs)$,
 we use $\varphi(x)\,\psi(T_{\frac{x}{\eta}}\tilde\omega)$ as test function in order to express the two-scale convergence 
 $\eta\nabla u_\eta\twoscale P_0$.  We obtain:
\begin{equation} \label{relD0}
\int_{ {B_R}\times \Omega} P_0(x,\omega)\,\varphi(x)\psi(\omega)\, dx\,d\PPt(\omega)=\lim_{\eta\to0}\, \eta\,\int_{{B_R}}\nabla u_\eta \cdot\varphi(x)\,\psi(T_{\frac{x}{\eta}}\tilde\omega)\, dx\ .
\end{equation}
Let assume that $(\varphi,\psi)$ is supported in $\mat\times\Sigma^*$. It follows from \eqref{eq.xinDeta} that the integration domain in the right hand side of (\ref{relD0}) reduces to $\mat\setminus\DomD_\eta$. The convergence $1_{{B_R}\setminus\DomD_\eta}\nabla u_\eta\twoscale \chi_0$ implies that 
the right hand side of (\ref{relD0}) vanishes as $\eta\to0$. First part of claim \eqref{eq.D0} follows by a trivial localization argument. The second part is deduce by the same kind of argument after
notice that $\nabla u_\eta$ remains bounded in $L^2(B_R\setminus\mat)$.

\noindent
In order to prove  (\ref{eq.nablau0}), we integrate by parts the right hand member of (\ref{relD0}). We obtain
$$
\eta\,\int_{{B_R}}\hspace{-0.2cm}\nabla u_\eta \cdot\varphi(x)\,\psi(T_{\frac{x}{\eta}}\tilde\omega)\, dx=-
\int_{{B_R}}\hspace{-0.2cm}u_\eta \Big(\eta\,\nabla\varphi(x)\cdot\psi(T_{\frac{x}{\eta}}\tilde\omega)+\varphi(x)\,(\Div_s\psi)(T_{\frac{x}{\eta}}\tilde\omega)\,\Big)\, dx\ 
$$
where the stochastic divergence appears by applying the chain rule formula \eqref{eq.chainrule}.
Passing to the limit $\eta\to0$, by using \eqref{relD0} and the convergence $u_\eta \twoscale u_0$, we deduce that
$$  \int_{{B_R}\times \Omega} P_0(x,\omega)\,\varphi(x)\psi(\omega)\, dx\,d\PPt(\omega)= - \int_{{B_R}\times\Omega} u_0(x,\omega) \, \varphi(x)\, \Div_s\psi(\omega)\, dx\,d\PPt(\omega)\ .
$$
The claim \eqref{eq.nablau0} follows by exploiting the integration by parts formula \eqref{eq.ibp}.

\med
Recalling \eqref{eq.xinDeta}, claim \eqref{eq.chi02} is a straightforward consequence of the two-scale convergence \goodbreak $1_{{B_R}\setminus\DomD_\eta}\nabla u_\eta\twoscale \chi_0(x,\omega)$
evaluated  with test functions of the kind $\varphi(x)\,\psi(T_{\frac{x}{\eta}}\tilde\omega)$ being $\psi$ supported in $\Sigma$.

\medskip\noindent
Next we multiply \eqref{eq.eta} by function $\eta\,\varphi(x)\, \psi(T_{\frac{x}{\eta}}\tilde\omega)$
with $(\varphi,\psi)\in C_c^\infty({B_R})\times H^1_s(\Omega)$, and  integrate by parts 
with the help of chain rule formula \eqref{eq.chainrule}. It comes
$$ \int_{{B_R}} a_\eta \nabla u_\eta \cdot \left(  \varphi(x) \, \Grads \psi(T_{\frac{x}{\eta}}\tilde\omega) +
 \eta\, \psi(T_{\frac{x}{\eta}}\tilde\omega) \nabla \varphi   \right) \ =\  \eta \, k_0^2 \int_{{B_R}} u_\eta\, \varphi(x)\, \psi(T_{\frac{x}{\eta}}\tilde\omega)\ .
$$
Then noticing that $a_\eta$ is of order $\eta^2$ in $\DomD_\eta$, we infer from \eqref{eq.twosc} that
$$
\int_{{B_R}\times\Omega} \chi_0(x,\omega) \cdot \varphi(x)\, \Grads \psi(\omega) \, dx\, d\PPt(\omega) \ =\ 0\ .
$$
Relation \eqref{eq.chi0} follows by localizing in $x$ and by using once more \eqref{eq.ibp}.

\QED

%
%
%
%
%


\begin{proposition} \label{t.var}  Let $u_0$ be the two-scale limit in \eqref{eq.twosc}. Then $u_0\in L^2(B_R; H^1_s(\Omega))$
and there exits a suitable $u\in L^2(B_R)$ such that:
$$
\begin{array}{llll}
 i)\qquad&u_0(x,\omega)=u(x)\quad&\hspace{-2cm}\mbox{a.e. in}\quad ({\mat}\times\Sigma^*) \ \cup \big[(B_R\setminus\mat)\times\Omega\big]\\
ii)\quad& \Divs \left( \boldsymbol{\varepsilon}^{-1}_0(\omega) \Grads u_0(x,\omega)\right)+\, k_0^2\,u_0(x,\omega)=0\qquad&\mbox{ in}\quad \mat\times\Sigma.\\
\end{array}
$$
\end{proposition}

\begin{remark}\upshape
In view of the characterization \eqref{eq.ibp},  the equation in $ii)$ has to be understood in the following weak sense: for a.e. $ x\in \mat$, it holds for every $\psi\in H^1_s(\Omega)$ vanishing in $\Sigma^*$ 
\begin{equation}\label{weakform}
 \int_{\Omega}  \boldsymbol{\varepsilon}^{-1} _0(\omega)\Grads u_0(x,\omega)\cdot \Grads \psi(\omega)\, d\PPt \ =\  k_0^2\int_{\Omega} u_0(x,\omega)
\,\psi(\omega)\, d\PPt \ .
\end{equation}

\end{remark}

\proof
%
From equation \eqref{eq.nablau0}  we deduce that $\nabla^su_0$ is an element of $L^2(B_R\times \Omega)$
and therefore $u_0\in L^2(B_R; H^1_s(\Omega))$. 
In addition, by \eqref{eq.D0} , we have
 $\nabla^su_0(x,\cdot)=0$ in $({\mat}\times\Sigma^*) \ \cup \big((B_R\setminus\mat)\times\Omega\big).$
Then  Claim $i)$ is a consequence of the assertion iii) of Proposition \ref{p.diff_stoc}.
Next in order to prove Claim ii), we consider $\varphi\in C_c^\infty(\mat)$ and 
$\psi\in C^0(\Omega)$ running in a dense subset of the subspace $\{\psi \in H^1_s(\Omega): \psi=0\ {\rm in}\ \Sigma^*\}$.
Then we multiply equation  \eqref{eq.eta} (for $\omega=\tilde\omega$) by   test function $\varphi(x) \, \psi(T_{\frac x\eta} \tilde\omega)$
 and integrate over $\mat$. Then taking into account chain rule \eqref{eq.chainrule}, we are led to
\begin{multline*}
  \int_{\mat}a_\eta(x,\tilde\omega)\,\nabla u_\eta\cdot \Big(\nabla \varphi(x)\,\psi(T_{\frac{x}\eta} \o)+\frac{1}{\eta}\,\varphi(x)\,\nabla^s\psi(T_{\frac{x}\eta}\o)\Big)\, dx = k_0^2\int_\mat u_\eta\,\varphi(x)\,\psi(T_{\frac{x}\eta} \o)\,dx\,  d\PPt.
\end{multline*}
In view of (\ref{eq.a_eta_fin}) and since $\psi$ vanishes outside $\Sigma$,  we may
 substitute $a_\eta(x,\tilde\omega)$ with $\eta^2 \left(\bs{\eps}_0(T_{\frac{x}{\eta}} \o)\right)^{-1}$ in the equality above. 
 Then we may pass to the limit $\eta\to 0$  by using the weak two-scale convergences in (\ref{eq.twosc}) and the product rule (\ref{product}):
\[
 \int_{\mat\times\Omega} \boldsymbol{\varepsilon}^{-1} _0(\omega) \, P_0(x,\omega)\cdot \Grads \psi(\omega)\, \varphi(x)\,dx\, d\PPt \ =\ k_0^2\int_{\mat\times\Omega}u_0(x,\omega)\,\varphi(x)\,\psi(\omega)\,dx\, d\PPt. \]
 Eventually, since $P_0=\nabla^s u_0$ (by \eqref{eq.nablau0}) and by the arbitrariness of function $\varphi$, 
we are led to relation \eqref{weakform} holding for a.a. $x\in \mat$ and for all $\Psi\in H^1_s(\Omega)$ vanishing outside $\Sigma$.
 \QED

\subsection{Identification of $\mathbf{u_0(x,\cdot)}$ }

The two-scale limit $u_0$  in Proposition \ref{t.var} can be identified as a solution of the Helmholtz resonator problem in a disk (see Lemma \ref{micro}).
It turns out, for a.e. $x\in \mat$, $u_0(x,\cdot)$ that it coincides up to a multiplicative factor with the random function $\Lambda$ introduced in
 \eqref{eq.lambda}.
 

\begin{proposition} \label{t.rappru} Assume that condition (\ref{eq.hyp_im}) holds. Then 
for a.e. $(x,\omega)\in B_R\times\Omega$ 
 \begin{equation}\label{decomp.u0}
u_0(x,\omega)=
\begin{cases} 
u(x) & \mbox{ in } (B_R\setminus\mat)\times\Omega, \\
u(x)\, \Lambda(\omega) & \mbox{ in }\mat\times\Omega.
\end{cases}
\end{equation}

\end{proposition}
%
%
%

\proof By the first assertion of Proposition \ref{t.var}, we know that $u_0(x,\cdot)= u(x)$ for $x\in B_R\setminus \mat$
whereas, for $x\in\mat$,  $u_0(x,\cdot) $ agrees with $u(x)$ on $\Sigma^*$ and satisfies equation \eqref{weakform} on $\Sigma$.
By using  the claim $i)$ of Proposition \ref{p.diff_stoc} and recalling that $\Sigma = \{ (y,m) : y\in B(\btheta_0(m),\brho_0(m))\}$, 
 we deduce that for a.e. $(x,m)\in \mat\times \Pi$, the function $u_0(x,m,\cdot)$
belongs to  $W^{1,2}(Y)$ and solves the boundary value problem
$$\begin{cases}
 \Delta_y \f + \, k_0^2 \,\bepsilon_0(m) \, \f = 0 \quad &\text{ in\quad $B(\btheta_0(m),\brho_0(m))$},\\
 \f= u(x) \quad& \text{ on\quad $\partial B(\btheta_0(m),\brho_0(m))$}\ .
\end{cases}$$
Thus in view of Lemma \ref{micro} and by linearity, the solution $\f$ is unique
 and reads  $ \f = u(x) \, w(y- \btheta_0(m))$, where  $w$ is given by the series \eqref{basicresonator} 
 whith $\rho=\brho_0(m)$ and  $\alpha:= k_0^2 \,\bepsilon_0(m)$.
 Indeed the assumption (\ref{eq.hyp_im})  ensures that $ \alpha\, (\brho_0(m))^2$ does not belong to $\bsigma$ (spectrum of Laplace operator on the unit disk). 
 Eventually, owing to  \eqref{eq.lambda}, one checks that  $\Lambda (m, \cdot ) = w(\cdot- \btheta_0(m))$ for $\pi$ a.e. $m\in  \Pi$.
 \QED

\begin{remark} 
\upshape
As a consequence of Propositions \ref{t.var} we know that $u_0(x,\cdot)\in H^1_s(\Omega)$ for a.e. $x\in \mat$. Thus
 relation \eqref{decomp.u0} implies  $\Lambda\in H^1_s(\spPt)$ unless $u\equiv 0$. 
To conclude this,  two hypothesis are needed:
\begin{itemize}
 \item   The condition (\ref{eq.hyp_im}), saying that \quad$\disp\p\big(\big\{(\theta,\rho,\eps)\in M\ ,\ \Im{m}(\eps)>0 \big\}\big)>0$,
 \item   The two-scale convergence \eqref{eq.twosc}; which will be in fact equivalent, up to a subsequence, to the $L^2$ bound \eqref{eq.bound} (see Lemma \ref{p.apriori}).
\end{itemize}
Accordingly, the property  $\Lambda\in H^1_s(\spPt)$ appears to be a \textit{necessary} condition in order to have \eqref{eq.bound}.
Actually, we will have to assume conditions \eqref{eq.hyp}\eqref{eq.hyp_im} in order to ensure that \eqref{eq.bound} holds for $\PPt$-almost all $\tilde\omega\in \tilde\spPt$
(see Proposition \ref{t.strongconv}).  As can be seen in  Lemma \ref{lem.Lambda} below, 
the fact that  $\Lambda$ belongs to $H^1_s(\spPt)$ can be derived directly from the assumption \eqref{eq.hyp} 
(in fact we need it merely for $r=0$)
\end{remark}

\begin{lemma}\label{lem.Lambda} 
 Under condition (\ref{eq.hyp}), we have $\Lambda\in H^1_s(Y)$ and it holds for every $ \o\in \O$:
\begin{equation}   \label{H1mat} 
\begin{split}
   \lim_{\eta\to0^+}\int_\mat |\Lambda(T_{\frac{x}{\eta}}\tilde\omega)|^2dx &= |\mat|\int_\Omega |\Lambda(\omega)|^2\PPt(d\omega),\\
   \lim_{\eta\to0^+}\int_\mat |\Grads\Lambda(T_{\frac{x}{\eta}}\tilde\omega)|^2dx &= |\mat|\int_\Omega |\Grads \Lambda(\omega)|^2\PPt(d\omega).
 \end{split}
\end{equation} 

\end{lemma}
 \proof 
 We notice that $\Lambda(m,\cdot)= 1$ near $\partial Y$  and 
 by Remark \ref{periodic}, in order to show that $\Lambda\in H^1_s(\Omega)$, we have only to check that $\Lambda\in L^2(\spP,\PP;W^{1,2}(Y))$.
 Recalling that $\int_Y \f_n \f_m\, dy= \delta_{nm}$ and $\int_Y|\nabla \varphi_n|^2\, dy=\lambda_n$, we can compute
\begin{equation} \label{e.lambdainH}
\|\Lambda-1\|^2_{L^2(\Omega)}=\sum_{n\in I_0} \alpha_n\, c_n^2
\ ,\qquad
\|\nabla_y\Lambda\|^2_{L^2(\Omega)}=\sum_{n\in I_0}\beta_n\, \lambda_n \, c_n^2,
\end{equation}
 where we denote
$$c_n := \int_{B_1}\varphi_n dy,  
\quad \alpha_n := \int_{M}\Big(\frac{k_0^2\varepsilon \rho ^3}{\lambda_n-k_0^2\varepsilon \rho ^2}\Big)^2d\p, \quad  
 \beta_n := \int_{M}\Big(\frac{k_0^2\varepsilon \rho ^2}{\lambda_n-k_0^2\varepsilon \rho ^2}\Big)^2d\p .$$
 From condition (\ref{eq.hyp}) (which clearly holds also if $r=0$) and the  fact that $\p$ is compactly supported,  
 we infer that $\sup_n \alpha_n<+\infty$. Furthermore, since $\lambda_n\to \infty$ as $n\to\infty$,
 there exists a constant $C>0$ such that for $n$ sufficiently large 
 $$ \lambda_n^2\,\beta_n \ =\int_M  \left( \frac{k_0^2\eps\rho^2}{1-\frac{k_0^2\eps\rho^2}{\lambda_n}}\right)^2\leq C\ .$$
 Thus, as $\sum_n |c_n|^2 <+\infty$ and  in view of equalities \eqref{e.lambdainH}, we deduce that $\Lambda\in H^1_s(\Omega)$
 with $\nabla^s \Lambda(m, \cdot)= \nabla_y \Lambda(m,\cdot)$ (see \eqref{e.derivative}).

\med
 Let us prove now \eqref{H1mat}. By applying Lemma \ref{p.strcont} and the chain rule 
 \eqref{eq.chainrule}, we find that the maps  $x\mapsto\Lambda(T_{\frac{x}{\eta}}\tilde\omega)$ and  $x\mapsto \nabla^s\Lambda(T_{\frac{x}{\eta}}\tilde\omega)$
 are in $L^2(\mat)$ and $(L^2(\mat))^2$ respectively.
Accordingly, by \eqref{ergomean},   we infer that \eqref{H1mat}
 holds for every  $\omega\in \O$.
 \QED


\subsection{Identification of vector field $\bs{\chi}_0$}
Let us point out that $\chi_0(x,\omega)$ defined by \eqref{eq.twosc} is associated up to a $\pi/2$ rotation to the  (transverse) electric field which is ocillating at scale $\eta$ in the obstacle $\mat$. 
In the following we will use the average of $\chi_0(x,\cdot)$ that is 
\begin{equation}\label{j}
j(x) \ =\ (j_1,j_2) =  \E(\chi_0) 
\end{equation}
We are going to determine  explicitely $\chi_0$ and $j$ in term of the effective permittivity tensor $\epsef$ given in (\ref{eq.epsef}) and 
the local gradient $\nabla u(x)$ of  function $u$ defined in \eqref{decomp.u0} (which, roughly speaking, describes 
the slowly varying magnetic field ``outside'' the rods).

\begin{proposition}\label{t.Aeff}\ Let $\bf{A}$ be the tensor defined in \eqref{globalepsmu}. Then
\begin{enumerate}
\item [i)]\ We have the relation:
\begin{equation}\label{decompchi}
\chi_0(x,\omega) =
\begin{cases}
j(x) &\text{in } (B_R\setminus\mat)\times \Omega\\
j_1(x)\,\sigma^1(\omega) + j_2(x)\,\sigma^2(\omega) &\text{in }\mat\times \Omega\ ,
\end{cases}
\end{equation}
 where for $i\in\{1,2\}$, $\sigma^i\in L^2_{sol}(\Omega)$ is the unique solution of
 \eqref{eq.epsef} with $\E(\sigma)=e_i.$

\item [ii)]\ $u$ belongs to $H^1(B_R)$ and the following constitutive relation holds
\begin{equation}\label{constitutive}
j(x) = \bf{A}(x)\,\nabla u=
\begin{cases}
\nabla u (x)  &\text{if  $x\in B_R\setminus\mat$}\\
( \epsef )^{-1}\, \nabla u(x)  &\text{if $x\in \mat$ }
\end{cases}
\end{equation}

\end{enumerate}
\end{proposition}



\proof

Let $\varphi \in C_c^\infty(B_R)$ and  $\bs{\psi}$ in a dense subset of $L^2_{\rm sol}(\spPt)$ such that $\bs{\psi}=0$  on $\Sigma$.
By \eqref{eq.xinDeta},  the function $x\mapsto \bs{\psi}(T_{\frac{x}\eta}\tilde\omega)$ vanishes in $\DomD_\eta(\tilde\omega)$ whereas, by Corollary \ref{stochdiv}, the vector field $x\mapsto {\psi}(T_{\frac{x}{\eta}}\tilde\omega)$ is divergence free. Thus:
\begin{multline*}
\int_{B_R}  \Big[\nabla_x u_\eta(x,\tilde\omega)1_{B_R\setminus \DomD_\eta(\o)}(x)\Big]\cdot \varphi(x)\,\bs{\psi}(T_{\frac{x}{\eta}}\tilde\omega)\,dx\\=
\int_{B_R } \nabla_x u_\eta(x,\tilde\omega)\cdot \varphi(x)\,\bs{\psi}(T_{\frac{x}{\eta}}\tilde\omega)\,dx
=-\int_{B_R} u_\eta(x,\o)\Big[\nabla\varphi(x)\cdot\bs{\psi}(T_{\frac{x}{\eta}}\tilde\omega)\Big]\,dx\ ,
\end{multline*}
Let us pass to the limit $\eta\to 0$. In view of the two-scale convergences in \eqref{eq.twosc} and 
noticing that $u_0(x,\cdot)\,\bs{\psi}=u(x)\,\bs{\psi}$ (see assertion $i)$ of Proposition \ref{t.var}), we obtain
\begin{eqnarray} \nonumber
\int_{B_R \times \spPt} \chi_0(x,\omega)\cdot\varphi(x)\,\bs{\psi}(\omega)\,dx\, d\PPt&=& -\int_{B_R\times \spPt} u_0(x,\omega)\nabla_x\varphi(x)\cdot\bs{\psi}(\omega)\,dx\, d\PPt\\  \label{chi0j0proof}
&=&-\int_{B_R} u(x)\nabla_x\varphi(x)\,dx\cdot\int_\Omega\bs{\psi}(\omega)\, d\PPt
\end{eqnarray}
Choosing first $\varphi$ to be compactly supported in $\mat$ and by localization, we deduce from \eqref{chi0j0proof} that
for a.e. $x\in \mat$, the vector field $\chi_0(x,\cdot)$ satisfies the equation
$$
\E (\chi_0(x,\cdot)\cdot\,\bs{\psi}) =0  \qquad \text{for all} \ \bpsi\in L^2_{\rm sol}(\Omega)\ \text{such that }\ \bpsi=0\ \text{in }\Sigma\ \text{ and } \E(\bpsi)=0\ .
$$
Recalling that $\chi_0(x,\cdot)$ vanishes on $\Sigma$ (see \eqref{eq.chi02}) and in view of \eqref{j}, we find that $ \chi_0(x,\cdot)$ satisfies the variational characterization
of the  solution of problem \eqref{eq.epsef} for $z=j(x)$.  Accordingly $\chi_0(x,\cdot)$ 
agrees for  $x\in \mat$  with the linear decomposition appearing in \eqref{decompchi}. 
 
 \med In order to prove that $\chi_0(x,\cdot)=j(x)$ for $x\in B_R\setminus \mat$, we take now $\varphi$ to be compactly supported in  $B_R\setminus \ov{\mat}$
and observe that relation \eqref{chi0j0proof} holds true for all $\bpsi\in L^2_{\rm sol}(\Omega)$  (it is not necessary that $\bpsi$ vanishes in $\Sigma$).
By localization, we find that, for a.e. $x\in B_R\setminus \mat$, it holds $\E (\chi_0(x,\cdot)\cdot\,\bs{\psi}) =0$ for $\bpsi\in L^2_{\rm sol}(\Omega)$
with vanishing average. As constant vector fields belong to $L^2_{\rm sol}(\Omega)$, we may choose in particular $\bpsi_0 = \chi_0(x,\cdot)- j(x)$ so that
$$ \E\left( |\chi_0(x,\cdot)- j(x)|^2\right)  \ =\ \E\left(\chi_0(x,\cdot)\cdot \bpsi_0 \right)- j(x)\cdot \E( \bpsi_0) \ =\ 0\ .$$
This ends the proof of assertion i). Let us prove now the assertion ii). To that aim, let us apply \eqref{chi0j0proof} with
$\bpsi= \sigma^i$ for $i=1,2$. Recalling that $\E(\sigma^i) = e_i$, we get
$$ \int_{B_R \times \spPt} \chi_0(x,\omega)\cdot\varphi(x)\, \sigma^i(\omega)\,dx\, d\PPt
\ =\ -\int_{B_R} u(x) {\partial\varphi \over \partial x_i}\,dx\ ,$$
holding for every $\varphi\in C^\infty_c(B_R)$. This allow to identify the distributional derivative of $u$ on $B_R$ as
\begin{equation}
{\partial u\over \partial x_i}\ =\ \E\left( \chi_0(x,\cdot) \cdot \sigma^i \right) \quad \ , \ i\in\{1,2\} \ .\label{u-derivative}
\end{equation}
Clearly, by Cauchy-Schwartz inequality, the right hand member of \eqref{u-derivative} is a function in $L^2(B_R)$.
Moreover, by applying  \eqref{decompchi}, we obtain that $\nabla u(x)= j(x)$ for a.e $x\in B_R\setminus\mat$ whereas 
$$ {\partial u\over \partial x_i}\ =\    \E(\sigma^1\!\cdot\!\sigma^i)\ j_1 +  \E(\sigma^2\!\cdot\!\sigma^i)\ j_2 \quad \text{ a.e. in $\mat$}\ . $$ 
We can then conclude that  relation \eqref{constitutive} holds in $\mat$ by using  the characterization \eqref{epstensor} of tensor $\epsef$. \QED

\section{Proof of the main homogenization result} \label{s.hom.prob}

From now on we consider a typical realization $\tilde \omega\in\O$ and study the asymptotic as $\eta\to 0$ of the solution $u_\eta:=u_\eta(x,\tilde\omega)$ of problem \eqref{eq.eta}.
We fix a reference ball $B_R$ of radius $R$ such that $\mat \Subset B_R$ is compactly embedded and
we admit for the moment the following upper bound estimate:
\begin{equation} \label{eq.bound}
\sup_{\eta>0} \int_{B_R}|u_\eta(x,\o)|^2dx <\infty.
\end{equation}
The validity of this estimate will be proved later under 
a specific assumption on the propbality $\p$.
Thanks to (\ref{eq.bound}), we can extract a subsequence of $u_\eta$ two-scale converging to some $u_0(x,\omega,\tilde \omega)$ in $L^2(B_R,dx)$.
Notice that \textit{a priori} the limit $u_0$ depends on the initial configuration $\tilde \omega$ of the obstacle and on the extracted subsequence.
As seen in Proposition \ref{t.rappru}, the limit does not depends neither of $\tilde \omega$ (that is $u_0=u_0(x,\omega)$) neither of the subsequence.

According with these remarks and to simplify notations, we shall denote by $u_0(x,\omega)$ the stochastic two-scale limit of $u_\eta(\cdot,\o)$,
and we do not recall every times its dependence with respect to $\tilde\omega$ and to the extracted subsequence.

The rest of this section is dedicated to the proof of the main result given in Theorem \ref{main}. We follow the next plan:

\med  {\em Step 1:\ } First we establish the behavior of the sequence $u_\eta(\cdot,\o)$ far from the obstacle $\mat$ (see Proposition  \ref{p.exterieur}).

\med  {\em Step 2:\ }In a second step, we assume  the bound (\ref{eq.bound}) and  we show that, under suitable conditions on the probability $\p$, we can pass
to the limit in \eqref{eq.eta} obtaining the homogenized problem \eqref{e.limiteq}. 
%
%

\med  {\em Step 3:\ }  In this last step, we improve the convergence obtained in Step 2 by proving the  \textit{strong} two-scale convergence of
the solutions (see Proposition \ref{forte}).  Then
 we deduce that, under conditions \eqref{eq.hyp} and \eqref{eq.hyp_im}, 
the bound \eqref{eq.bound} holds for $\PPt$-almost all $\tilde \omega\in \O$,
 (in fact for all $\o\in \O$ such that relations in Lemma \ref{lem.strong.conv} are satisfied).

\med As a  conclusion we will obtain the homogenization result under the sole  assumptions \eqref{eq.hyp} \eqref{eq.hyp_im}
on the probability distribution $\p$ of the dielectric rods.

%

\subsection{Behavior far of the obstacle}

This first step is related to the convergence of $u_\eta$ at a positive distance from $\mat$.
We use the following result whose proof relies on the hypo-ellipticity of operator $\Delta + k_0^2$
and is very similar to that used  in \cite[Lemma 2.1]{BF06} in the context of a 3D-diffraction problem.   
\begin{proposition}\label{p.exterieur}
Fix $\tilde\omega\in\Omega$ and let $u_\eta(\cdot,\tilde\omega)$ the solution of problem (\ref{eq.eta}) with outgoing Sommerfeld radiation condition
for an incident wave $u^i_\eta$.
Let us assume that $u^i_\eta$ converges uniformly to $u^i$
and that $u_\eta\rightharpoonup \mathbf{u}$ weakly in $L^2(B_R)$.
Then the convergence $u_\eta\to \mathbf{u}$ holds in $C^\infty(K)$ for any compact set $K\subset \Rset^2\setminus \overline\mat$.
Furthermore, the limit field $\mathbf{u}$ satisfies the Helmholtz equation $\Delta \mathbf{u} +k_0^2\, \mathbf{u}=0$ in $\Rset^2\setminus\overline\mat$ and $(\mathbf{u}-u^i)$ satisfies the outgoing Sommerfeld radiation condition.
\end{proposition}



\med We can now  derive an important additional energy estimate: 
\begin{lemma} \label{p.apriori} 
Let $\tilde\omega\in\tilde\Omega$ such that condition (\ref{eq.bound}) holds. Then it holds
$$\sup_{\eta} \int_{B_R} \, a_\eta(x,\tilde\omega)|\nabla_x u_\eta(x,\tilde\omega)|^2 \, dx \ <\ +\infty.$$
As a consequence the sequences  $(\eta\nabla u_\eta(x,\tilde\omega))$ and $(1_{B_R\setminus\DomD_\eta(\o)}(x)\nabla u_\eta(x,\tilde\omega))$ are bounded in $L^2(B_R)$.
\end{lemma}
\proof  Recalling that $a_\eta=1$ in $B_R \setminus \mat$,
we multiply equation \eqref{eq.eta} by ${\ov u}_\eta$ and integrate by parts on  $B_R$, obtaining
\begin{multline*}
 -\int_{B_R} a_\eta(x,\tilde\omega)|\nabla_x u_\eta(x,\tilde\omega)|^2 \, dx+ k^2_0\,  \int_{B_R} |u_\eta(x,\tilde\omega)|^2\, dx 
=\int_{\partial B_R} u_\eta(x,\tilde\omega)\frac{\partial \overline{u_\eta}}{\partial n}(x,\tilde\omega)dx\ .
\end{multline*}
Thanks to Proposition \ref{p.exterieur}, the right-hand side of the previous relation is uniformly bounded. Combined with
 (\ref{eq.bound}) we obtain the desired upper bound. The last assertion follows from the definition of $a_\eta$ in \eqref{eq.a_eta_ini}.
 \QED

\subsection {Derivation of the homogenized problem}

As pointed out in Section \ref{s.mainresult}, the limit problem is characterized by two functions
\[
  \mathbf{A}(x) =1_{\Rset^2\setminus\mat}(x) + 1_{\mat}(x)\, (\epsef)^{-1},\quad   \bs{b}(x,k_0)=1_{\Rset^2\setminus\mat}(x)+\mu^{\mathrm{eff}}(k_0)1_{\mat}(x),
\]
with $\epsef, \muef$ given in \eqref{eq.epsef} and \eqref{def.mueff} respectively.
Then we recall the  \textit{homogenized problem} given in (\ref{e.limiteq}) 
\begin{equation*}
 \begin{cases}
    \Div\big( \bs{A}(x)\nabla u(x)\big)+k_0^2\,\bs{b}(x,k_0)\, u(x)=0,\qquad x\in \Rset^2,\\
     u-u^{\rm inc}\quad \text{\upshape satisfies the outgoing Sommerfeld radiating condition}.
  \end{cases}
\end{equation*}

\begin{lemma}[uniqueness] \label{unique} Assume that $\Im(\muef)>0$. Then the solution to (\ref{e.limiteq}) is unique in $W^{1,2}_{\rm loc}(\R^2)$.
\end{lemma}

\proof Let $u\in W^{1,2}_{\rm loc}$ be solving (\ref{e.limiteq}) for $u^{\rm inc}=0$. By multiplying the first equation by $\ov u$ and integrating by parts over $B_R$,
we obtain:
$$ \int_{B_R} \left(\bs{A}\nabla u \cdot \nabla\ov u - k_0^2 \, \bs{b} |u|^2\right) \, dx = \int_{\partial B_R} \ov u \, \frac{\partial u}{\partial n} 
\ .$$
As $u$ satisfies the outgoing wave condition \eqref{SM}, following classical arguments (see \cite{Co}), it is easy to infer that the right hand member in the equality above 
 has a non negative imaginary part. Exploiting that $\bs{A}$ is positive definite whereas $\Im(\muef)>0$, we deduce that $u$ vanishes a.e. on $\mat$.
 Therefore by the transmission conditions in \eqref{transmission}, it holds $ u^+= \partial_n u^+ =0$ on $\partial \mat$. As $u$ solves $(\Delta + k_0^2) u=0$ on $\R^2\setminusÊ\ov\mat$, it is straightforward that $u$ vanishes over all $\R^2$. 
 \QED

 \med
 
 \med
 Our second step consists in proving the following convergence result:

\begin{proposition}\label{t.lim_prob}
Let us assume (\ref{eq.hyp}) (\ref{eq.hyp_im}) and let $\o\in\O$ such that condition (\ref{eq.bound}) is satisfied.  
Then the sequence of solutions $u_\eta(\cdot,\tilde\omega)$ converges weakly in $L^2_{\mathrm{\loc}}(\Rset^2)$ 
to $\bs{b}(\cdot)\,u(\cdot) $, where $u$ is  the unique solution 
of the homogenized problem \eqref{e.limiteq}.

\end{proposition}

\proof  By Lemma \ref{p.apriori} and the bound \eqref{eq.bound}, there exists  $u_0(x,\omega)\in L^2(\mat\times \spPt, \mathcal L^2\otimes \PPt)$ and
vector fields  $\chi_0(x,\omega)$, $P_0(x,\omega)$ in $ \big(L^2(\mat\times \spPt,dx\otimes \PPt)\big)^2$ such that the two-scale convergences \eqref{eq.twosc} hold up to a subsequence. In particular for such a subsequence (still denoted with the same symbol), we have 
$$u_\eta(x,\tilde\omega)\twoscale u_0(x,\omega)\quad ,\quad a_\eta(x,\tilde\omega)\nabla_x u_\eta(x, \tilde\omega)\twoscale \chi_0(x,\omega)\ ,$$
where $u_0$ is given by  Proposition \ref{t.rappru} whereas $\chi_0(x,\cdot)$ is given by Proposition \ref{t.Aeff}. We deduce 
the following weak convergences in $L^2(B_R)$:
\begin{equation} \label{e.weak_a}
\begin{split}
u_\eta(x, \tilde\omega)&\rightharpoonup\E(u_0(x,\cdot))=\bs{b}(x)u(x)\ ,\\
 a_\eta(x,\tilde\omega)\nabla_x u_\eta(x, \tilde\omega) &\rightharpoonup\E(\chi_0(x,\cdot))= \mathbf{A}(x)\nabla u(x)\, .
\end{split}
\end{equation}
Then we may pass to the limit in the distributional sense in (\ref{eq.eta}) so that
  $u$ solves the first equation of \eqref{e.limiteq} in $B_R$.
The fact that $u$ solves Helmholtz equation in $\R^2\setminus \mat$  with $(u-u^{\rm inc})$ satisfying the outgoing Sommerfeld condition \eqref{SM} is a consequence of Proposition \ref{p.exterieur}. Summarizing we have proved that $u$ as an element of $W^{1,2}_{\rm loc}$ solves the system \eqref{e.limiteq}. Thanks to  \eqref{eq.hyp_im}, we have $\Im(\muef)>0$
so that this limit $u$ is unique by Lemma \ref{unique}. In particular the   
 whole sequence $u_\eta(\cdot , \o)$ does converge weakly to $\bs{b}(\cdot) u(\cdot)$ in $L^2_{\rm loc}(\Rset^2)$ as stated in the Proposition.
\QED


%

%

\subsection{Strong two-scale convergence and proof of the $L^2$-bound}

%
%
\noindent
We start with a technical lemma where we use the discrete spectrum $\bsigma$ of the Dirichlet Laplace operator on the unit disk and the random positive numbers
\begin{equation*}\label{defX}
X_j(\omega) =  \frac{1}{\dist (\beps_j(\omega)\, \brho_j^2(\omega)\,  k_0^2,\bsigma)}\ .
\end{equation*}

\begin{lemma}\label{lem.strong.conv}
 Under condition (\ref{eq.hyp}), we have for almost every $ \omega\in\Omega$ :
\begin{equation}   \label{eq.strong.conv} 
 \limsup_{\eta\to0^+}  \Big( \eta^2 \, \sum_{j\in J_\eta(\omega)}  [X_j(\omega)]^{2+r}\Big) \, <\infty \quad ,\quad \lim_{\eta\to0^+} \Big(\eta^2\, \sup_{j\in J_\eta(\omega)}   [X_j(\omega)]^{2+r}\Big) \ =\ 0
\end{equation} 
\end{lemma}
 \proof
We notice that $\sharp (J_\eta(\omega))\sim|\mat| \eta^{-2}$. Then as the random variables $X_j(\omega)^{2+r}$ are independent and identically distributed, we may apply the strong law of large numbers. Thus
 for almost all $\omega\in\Omega$ we have
\begin{multline*}
 \lim_{\eta\to 0}\eta^2  \sum_{j\in J_\eta(\omega)} [ X_j(\omega)]^{2+r} =
\lim_{\eta\to 0} \frac{|\mat|}{\sharp J_\eta(\omega)}\sum_{j\in J_\eta(\omega)}[ X_j(\omega)]^{2+r}
= |\mat|\int_M[ X_j(\omega)]^{2+r}\,d\mathbb{P}(\omega)\ ,
\end{multline*}
where the right-hand side integral is finite by  hypothesis (\ref{eq.hyp}).
For the second part of \eqref{eq.strong.conv}, it is enough to apply Lemma \ref{l.sup} to the sequence $[X_j(\omega)]^{2+r} $.
\QED

\begin{proposition} \label{forte}
 Under the conditions of Proposition \ref{t.lim_prob} and assuming in addition that $\o$ is such that relations \eqref{eq.strong.conv}  hold,  
then, for any ball $B_R \Supset \mat$, we have
\begin{equation}\label{eq.stron_two_sacle}
   \lim_{\eta\to0^+}\int_{B_R}\left| u_\eta(x,\tilde\omega) - u_0(x,T_{\frac{x}{\eta}}\tilde\omega)\right|^2dx =0\ ,
\end{equation}
being $u_0$  given by \eqref{decomp.u0}.
Consequently $u_\eta(\cdot,\tilde\omega)$  strongly two-scale converges  to $u_0$. \end{proposition}

%
%
%
%
%
The proof of the Proposition follows directly from Lemmas \ref{l.conv1} and \ref{l.conv2} below where 
the convergence in $L^2$ norm is established on subsets $B_R\setminus \DomD_\eta$  and  $\DomD_\eta$ respectively.
In order to lighten the presentation, we use the following notations 
\begin{equation*}
 Y_{j,\eta}=\eta(j-\by(\o)+Y),\quad  \DomD_{j,\eta}=\eta(j-\by(\o)+B(\btheta_j(\o),\brho_j(\o)))\ ,\quad Y_{j,\eta}^*=Y_{j,\eta}\setminus \DomD_{j,\eta}
\end{equation*} 
that is  we prefer to write $Y_{j,\eta}$ instead of $Y_{j,\eta}(\tilde\omega)$ and so on for the other terms.

%
%
%
 
%
%
\begin{lemma}   \label{l.conv1}
 We have
\begin{equation}  \label{eq.convLp}
    \lim_{\eta\to0^+}\|u_\eta(\cdot,\tilde\omega)- u\|_{L^p(B_R\setminus \DomD_\eta)}=0, \quad \forall p\in [1,\infty).
\end{equation}
\end{lemma}

\proof
 We modify $u_\eta$ in the subset $\DomD_\eta(\o)$ by considering the harmonic extension. This way we obtain the new function $\tilde u_\eta\in H^1(B_R)$ defined by  
\[
   \begin{cases}
                   \tilde u_\eta(x)=u_\eta(x,\tilde\omega), & x\in B_R\setminus \DomD_\eta\\
                   \Delta \tilde u_\eta(x)= 0    & x\in \DomD_\eta\ ,\\
   \end{cases}
\]
By the rescaled inequality  \eqref{e.poincarre_rescal}, we have
\begin{align*}
  \| \tilde u_\eta\|_{H^1(B_R)}^2&=\|u_\eta(\cdot,\tilde\omega)\|_{H^1(B_R\setminus \DomD_\eta)}^2+ \sum_{i\in J_\eta} \|\tilde u_\eta \|_{H^1(\DomD_{j,\eta})}^2
\\
  &\hspace{-1cm}\leq \|  u_\eta(\cdot,\tilde\omega)\|_{H^1(B_R\setminus \DomD_\eta)}^2+ c_\delta \sum_{i\in J_\eta} \|\tilde u_\eta\|_{H^1(Y_{j,\eta}^*)}^2
   = (1+ c_\delta) \|  u_\eta(\cdot,\tilde\omega)\|_{H^1(B_R\setminus \DomD_\eta)}^2 \\
   & \le (1+ c_\delta)  \, \int_{B_R}  \left(|u_\eta(x,\o)|^2 + a_\eta(x,\o)\, |\nabla u_\eta(x,\o)|^2\right)\, dx\ .
\end{align*}
By \eqref{eq.bound} and the estimate in Lemma \ref{p.apriori}, the last integral is uniformly bounded.
Therefore, by the Sobolev compact embedding $H^1(B_R)\subset L^p(B_R)$ for $p\in [1,\infty)$,
 we can extract a convergent sequence (still denoted by $\tilde u_\eta$)  such that $\tilde u_\eta \to \tilde u$ in $L^p(B_R)$. 
It remains to check that  $\tilde u=u$. To that aim we observe that $\tilde u_\eta - u_\eta \twoscale \tilde u(x) - u_0(x,\omega)$ whereas 
we have the strong two-scale convergence $1_{B_R\setminus \DomD_\eta(\o)}(x)\stwoscale1_{B_R\times\Sigma^*}(x,y)$. Therefore by applying the product rule (\ref{product}), we deduce that
$$
  (\tilde u_\eta- u_\eta) \,1_{B_R\setminus \DomD_\eta(\o)}\ \twoscale \ (\tilde u(x) - u_0(x,\omega)) \, \,1_{B_R\times\Sigma^*}
$$
As the equalities $\tilde u_\eta(x)=u_\eta(x,\tilde\omega)$ and $u_0(x,\omega)= u(x)$ hold in  $B_R\setminus \DomD_\eta$, we conclude that $\tilde u=u$ a.e. in $B_R$.

\QED

\begin{lemma} \label{l.conv2}
 Let us set $v_\eta(x)=u_\eta(x,\tilde\omega) - u_0(x,T_{\frac{x}{\eta}}\tilde\omega)$. Then if $\tilde\omega$ satisfies (\ref{eq.strong.conv}), we have
\[
  \lim_{\eta\to0^+} \|v_\eta\|_{L^2(\DomD_\eta)} =0. 
\]
\end{lemma}

\proof
We write $\|v_\eta\|_{L^2(\DomD_\eta)}^2=\sum_{j\in J_\eta}\|v_\eta\|_{L^2(\DomD_{j,\eta})}^2 $ and 
apply, for every $j\in J_\eta$, the rescaled inequality (\ref{poincare2_rescal}) to each term $\|v_\eta\|_{L^2(\DomD_{j,\eta})}$ choosing $\alpha=\varepsilon_jk_0^2$. We get
%
%
\begin{multline}\label{6.7}
  \|v_\eta\|_{L^2(\DomD_\eta)}^2 \leq \sum_j\frac{2\,\rho_j^2\,\eta^4}{\mathrm{dist}^2(\eps_j\,\rho_j^2\,k_0^2,\bsigma) }\|\Delta v_\eta+\frac{\eps_j\, k_0^2}{\eta^2} v_\eta\|_{L^2(\DomD_{j,\eta})}^2
\\+4c_\delta\sum_j\left(\frac{|\eps_j|^2k_0^4\rho_j^4}{\mathrm{dist}^2(\eps_j\,k_0^2\,\rho_j^2,\bsigma) }+1 \right)
\left(\|v_\eta\|_{L^2(Y_{j,\eta}^*)}^2+ \eta^2\|\nabla v_\eta\|_{L^2(Y_{j,\eta}^*)}^2\right)\ .
\end{multline}
As $\tilde\omega$ satisfies (\ref{eq.strong.conv}), we have the following convergence
\begin{equation}\label{b_eta}
b_\eta\ :=\ \sup_{j\in J_\eta} \eta^2\, [X_j(\tilde\omega)]^2 \ \to \ 0. 
\end{equation}
On the other hand,  recalling that the $(\rho_j,\eps_j)$ remain in a bounded set of $\Rset^+\times \C$ and that $\varepsilon_j=\boldsymbol{\varepsilon}_0(T_{\frac{x}{\eta}}\tilde\omega)$ for $x\in \DomD_{j,\eta}$, we deduce from  \eqref{6.7} the  existence of a suitable constant $\kappa>0$ (depending only on $\delta$)  such that
 \begin{multline}\label{e.in_strong}
  \kappa\, \|v_\eta\|_{L^2(\DomD_\eta)}^2 \leq  (b_\eta+\eta^2)\|\nabla v_\eta\|^2_{L^2(B_R\setminus\DomD_\eta)}+ \|v_\eta\|_{L^2(B_R\setminus\DomD_\eta)}^2\\+
\sum_{i\in J_\eta} \Big( X_j^2(\omega)   \int_{Y_{j,\eta}^*}|v_\eta(x)|^2 \Big)
+ b_\eta\,\eta^2\, \left\|\Delta v_\eta+\frac{\beps_0(T_{\cdot/\eta}\,\o)\, k_0^2}{\eta^2} v_\eta\right\|_{L^2(\DomD_\eta)}^2\ .
\end{multline}
Therefore, in view of \eqref{b_eta} and \eqref{e.in_strong}, the Lemma is proved provided one shows the 
following claims
$$
\begin{array}{ll}
i)\quad \dis\|v_\eta\|_{L^2(B_R\setminus\DomD_\eta)}^2\to 0\\
ii)\quad \dis\sup_{\eta>0}\|\nabla v_\eta\|^2_{L^2(B_R\setminus\DomD_\eta)}<\infty\ ,\\
iii)\quad \dis\sup_{\eta>0}\, \eta^2\, \big\|\textstyle\Delta v_\eta+\frac{\beps_0\left(T_{\frac\cdot\eta}\,\o\right)\, k_0^2}{\eta^2} v_\eta\big\|_{L^2(\DomD_\eta)}^2<+\infty\ ,     \\
iv) \quad\dis  \sum_{j\in J_\eta} \Big( X_j^2(\omega)   \int_{Y_{j,\eta}^*}|v_\eta(x)|^2 \Big)\to 0 \ .
\end{array}
$$
Recalling that $T_{\frac{x}{\eta}}\tilde\omega\in \Sigma^*$ for $x\in B_R\setminus \DomD_\eta$ (see  \eqref{eq.xinDeta}) while
$u_0(x,T_{\frac{x}{\eta}}\tilde\omega)=u(x)$ for $x\in B_R\setminus\DomD_\eta$ (see Proposition \ref{t.var}, $i)$),
Claim $i)$ follows  directly from Lemma \ref{l.conv1}. On the other hand
as $\nabla v_\eta(x)= \nabla u_\eta(x,\tilde\omega)-\nabla u(x)$ holds in $B_R\setminus\DomD_\eta$, we deduce Claim $ii)$
taking into account that $u\in H^1(B_R)$ (see  Proposition \ref{t.Aeff}) and 
the gradient estimates in Lemma \ref{p.apriori}.

\med Let us prove now Claim $iii)$. As $u_\eta$ solves (\ref{main}), we get for $x\in\DomD_{j,\eta}(\o)$ :
\begin{equation}\label{lapveta}
\begin{split}
\Delta v_\eta(x)+\frac{\eps_j\, k_0^2} {\eta^2} v_\eta(x)&=\Delta_x u_\eta+a_\eta(x,\o)u_\eta-\Delta_x\Big(u_0(x,T_{\frac{x}\eta}\,\o)\Big)-
\frac{\eps_j\, k_0^2} {\eta^2} u_0(x,T_{\frac{x}\eta}\,\o)\\
&=-\Delta_x\Big(u_0(x,T_{\frac{x}\eta}\,\o)\Big)-\frac{\eps_j\, k_0^2} {\eta^2} u_0(x,T_{\frac{x}\eta}\,\o)\ .
\end{split}
\end{equation}
On the other hand, thanks to relation $u_0(\cdot,\omega)=u(\cdot)\Lambda(\omega)$ in $\mat$ and the chain rule (\ref{eq.chainrule}) it comes
\begin{equation}\label{lapu0}
 \Delta_x\Big(u_0(x,T_{\frac{x}\eta}\,\o)\Big)= \Delta_xu(x)\Lambda(T_{\frac{x}{\eta}}\tilde\omega)+\frac2\eta \nabla u(x)\cdot\nabla^s\Lambda(T_{\frac{x}{\eta}}\tilde\omega)+\frac1{\eta^2}u(x)\Delta_s\Lambda(T_{\frac{x}{\eta}}\tilde\omega)\ .
\end{equation}
Plugging (\ref{lapveta}) and (\ref{lapu0}) yields to
\begin{equation*}
\begin{split}
\Delta v_\eta(x)+\frac{\eps_j\, k_0^2} {\eta^2} v_\eta(x)&=
-\Delta_xu(x)\Lambda(T_{\frac{x}{\eta}}\tilde\omega)-\frac2\eta \nabla u(x)\cdot\nabla^s\Lambda(T_{\frac{x}{\eta}}\tilde\omega)\\
&-\frac{1}{\eta^2}\Big(\Delta_su_0(x,T_{\frac{x}{\eta}}\tilde\omega)-\eps_j\, k_0^2\, u_0(x,T_{\frac{x}\eta}\,\o)\Big)\ .
\end{split}
\end{equation*}
Recalling that $(T_{x/\eta}\o)\in\Sigma$ (see (\ref{eq.xinDeta})) and that $u_0$ solves the equation appearing in assertion
$ii)$ of Proposition \ref{t.var},
we deduce after multipling by $\eta$
\begin{equation}\label{e.stron_conv_demo}
\eta \, \left(\Delta v_\eta(x)+\frac{\eps_j\, k_0^2} {\eta^2} v_\eta(x)\right)\ =ᅵ
-\eta\, \Delta_xu(x)\Lambda(T_{\frac{x}{\eta}}\tilde\omega)-2\, \nabla u(x)\cdot\nabla^s\Lambda(T_{\frac{x}{\eta}}\tilde\omega)\ . \end{equation}
By Lemma \ref{lem.Lambda},  we have $\Lambda\in H^1_s(\Omega)$ from which follows that 
functions $x\mapsto \Delta_xu(x)\Lambda(T_{\frac{x}{\eta}}\tilde\omega)$ and $x\mapsto \nabla u(x)\cdot\nabla^s\Lambda(T_{\frac{x}{\eta}}\tilde\omega)$ 
are bounded in $L^2(\Omega)$ (and \textit{strongly two-scale converge} to $\Delta_xu(x)\Lambda(\omega)$ and $u(x)\cdot\nabla^s\Lambda(\omega)$ respectively). 
Thus Claim $iii)$ is a consequence of (\ref{e.stron_conv_demo}).

\med It remains to show Claim $iv)$.  Let $q=(2+r)/2$, being $r$ the parameter introduced in the hypothesis
\eqref{eq.hyp} and let  $q^*$ be such that $q^{-1}+{q^*}^{-1}=1 $.
By applying  H\"older inequality first on the sum over $j\in J_\eta$ and then on the integral, it comes
\begin{align*}
   \sum_{j\in J_\eta} X_j(\omega)^2\, \int_{Y_{j,\eta}^*(\omega)}|v_\eta|^2
    &\leq \left[\sum_{j\in J_\eta} X_j(\omega)^{2q} \right]^{\frac{1}q}
\left[\sum_{j\in J_\eta}\left(\int_{Y_{j,\eta}^*}|v_\eta|^2\right)^{q^*}\right]^{\frac{1}{q^*}}
\\
    &\leq\left(\sum_{j\in J_\eta(\omega)} \eta^2  X_j(\omega)^{2q}\right)^\frac1{q}
   \|v_\eta\|_{L^{2q^*}(B_R\setminus\DomD_\eta)}^2.
\end{align*}
By \eqref{eq.strong.conv} 
 the first factor on the right hand side remains bounded as $\eta\to 0^+$
while the second factor vanishes in virtue of Lemma \ref{l.conv1}.
This implies Claim $iv)$.  The proof of Lemma \ref{l.conv2} is complete.
 \QED
 
 \med
The last step consists in proving the energy bound by a contradiction argument.
\begin{proposition}[Energy bound] \label{t.strongconv}
Assume that \eqref{eq.hyp} and \eqref{eq.hyp_im} hold. Then the bound \eqref{eq.bound} holds $\PPt$-almost surely in any open ball $B_R$ such that  $\mat\Subset B_R$. 
\end{proposition}

\proof
 Let $\mathcal {Q}\subset\Omega$ be the set of all $\omega$ such that (\ref{eq.strong.conv}) holds.
 As by Lemma \ref{lem.strong.conv} $\mathcal {Q}$ is of full measure, it is enough to show that  
 the bound \eqref{eq.bound} holds true for every $\omega\in \mathcal {Q}$.
  Assume by contradiction that, for some $R>0$, \eqref{eq.bound} does not hold for some particular $\tilde\omega\in \mathcal {Q}$.
Then there exists a sequence $\eta_j\searrow 0$, $j\in \Nset$ such that
\[
  t_j= \|u_{\eta_j}(\cdot,\tilde\omega)\|_{L^2(B_R)}   \nearrow \infty\quad \text{as $j\to \infty$}.
\]
By the linearity of the problem, the normalized function $v_j= \frac{u_{\eta_j}(\cdot,\tilde\omega)}{t_j}$ satisfies
\begin{equation*} \label{eq.contrad}
\begin{cases}
  \Div \left(a_{\eta_j}(x,\tilde\omega)\nabla  v_j(x)\right) +k_0^2\, v_j(x)=0 \qquad x\in \Rset^2\\
  \|v_j\|_{L^2(B_R)}=1,\\
  v_j-\frac{u^{\mathrm inc}}{t_j}\text{ satisfies the Sommerfeld radiating condition.} 
\end{cases}
\end{equation*}
This means that $v_j$ solves the diffraction problem (\ref{eq.eta}) for $ \eta= \eta_j$ and
 a  incident wave $\frac{u^{\mathrm inc}}{t_j}$ tending to zero as $j\to\infty$.
  As by construction $\{v_j\, ;\, j\in \Nset\}$ is bounded in $L^2(B_R)$,
we may apply Proposition \ref{forte} and we are led to 
\begin{equation}\label{eq.energy_proof}
\lim_{j\to \infty} \int_{B_R} | v_j(x,\o)- v_0(x,T_{x/\eta}\o)|^2 \, dx \ =\ 0 \ ,\end{equation}
where $ v_0(x,\omega) =  v(x)[1_{\Rset^3\setminus \mat}(x)+\Lambda(\omega)1_{\mat}(x)]$ being $v$ the solution of (\ref{e.limiteq}) for a vanishing incident wave.
 Here the random function $\Lambda$ is the one defined in (\ref{eq.lambda}).
By the uniqueness Lemma \ref{unique}, we have $v\equiv 0$, thus $v_0\equiv 0$ which is uncompatible with  \eqref{eq.energy_proof} 
and the assumption $\|v_j\|=1$.
%
\QED

\section{Appendix}
In the following Lemma we keep  the notations of Section 2 for the spectral Dirichlet problem in 
the unit disk $D$ of $\Rset^2$. We denote $B_\rho=\rho D=B(0,\rho)$.
\begin{lemma}[Resonator problem] \label{micro} Let $\rho>0$ and $\alpha\in \C$ such that
$\rho^2\, \alpha \notin \bsigma$. Then 
\begin{enumerate}
 \item[i)] The boundary value problem $\Delta w + \alpha\, w =0 \ ,\ w=1 \ \text{on $\partial B_\rho$}$ 
 has a unique solution $w\in W^{1,2}(B_\rho)$ given by
\begin{equation}\label{basicresonator}
w(y) = 1 + \sum_{n=1}^{+\infty} \frac{\alpha \rho^2\, c_n}{\lambda_n-\alpha\,\rho^2} \, \varphi_n\Big(\frac{y}{\rho}\Big) \quad, \quad {\rm with}\ c_n=\int_D \varphi_n\  .
\end{equation}
 \item[ii)] For all $v\in W^{1,2}_0(B_\rho) \cap W^{2,2}(B_\rho)$, it holds 
$$ \Vert \Delta v+  \alpha v\Vert_{L^2(B_\rho)} \ \ge\ {\rm dist}(\alpha, \frac{\bs{\sigma}_0}{\rho^2}) \, \Vert  v\Vert_{L^2(B_\rho)}
 \ .
$$
\end{enumerate}

\end{lemma}
\proof  Straightforward by using the spectral decomposition of the Laplace operator 
on $B_\rho$ whose eigenvalues are $\{\frac{\lambda_n}{\rho^2} \}$ with normalized eigenvectors
$\{ \frac1{\sqrt{\rho}} \varphi_n(\frac{x}{\rho}) \}$.
\QED

\begin{lemma} \label{l.laplace}
Let  $\delta\in (0,1/2)$. Then there exists $c_\delta>0$ such that, for every $(\theta,\rho)$ with $\dist(\theta,\partial Y)\geq \rho+\delta$ and  for every $u\in W^{1,2}(Y)$ verifying
$\Delta u=0$ \ in $ B(\theta,\rho)$, there holds
\begin{equation}\label{e.poincarre2}
\int_{ B(\theta,\rho)}|\nabla u|^2\leq c_\delta\int_{Y\setminus B(\theta,\rho)}|\nabla u|^2.
\end{equation}
\begin{equation}\label{e.poincarre}
\int_{ B(\theta,\rho)}|u|^2\leq c_\delta\int_{Y\setminus B(\theta,\rho)}\Big(|u|^2+|\nabla u|^2\Big).
\end{equation}
\end{lemma}
\proof
In a first step we show the result assuming that  $B(\theta,\rho)$ is the largest ball in $Y$ such that $\dist(\theta,\partial Y)\geq \rho+\delta$ namely  $\ov B=B(\ov \theta,\frac{1}2-\delta)$ with  $\ov \theta= (\frac12,\frac12)$.

\med
Proof of (\ref{e.poincarre2}) : 
Assume by contradiction that (\ref{e.poincarre2}) does not hold.
Then there exists a sequence $(u_n)\subset W^{1,2}(Y)$ such that
\begin{equation}\label{e.contradiction}
\int_{\ov B}|\nabla u_n|^2=1\ ,\quad \int_{Y\setminus \ov B}|\nabla u_n|^2\to0\ ,\quad \Delta u_n=0 \mbox{ in } \ov B.
\end{equation}
Without any loss of generality we may assume additionally that $\int_Y u_n =0$. Thus, by Poincar\'e inequality,  $\{u_n\}$ is bounded  in $W^{1,2}(Y)$ and, possibly after extracting a subsequence, does converge  
 strongly in $L^2(Y)$  to some limit $u\in W^{1,2}(Y)$. From (\ref{e.contradiction})
 it comes that $\nabla u=0$ in $Y\setminus\ov B$ while $\Delta u=0$ in $B$. Therefore $u$ is constant and vanishes  by the zero average condition. In addition the convergence $u_n$ to $0$ over $Y\setminus \ov B$ is strong in  $W^{1,2}(Y\setminus \ov B)$ yielding in particular that 
the trace of $u_n$ does converge strongly to zero in $W^{1/2,2}(\partial B)$.
As $u_n$ is harmonic in $\ov B$, we deduce that $\int_{\ov B}|\nabla u_n|^2 \to 0$ in contradiction with (\ref{e.contradiction}).

\med
Proof of (\ref{e.poincarre}) : 
Assume by contradiction that (\ref{e.poincarre}) does not hold. Then there exists a sequence $(u_n)\subset W^{1,2}(Y)$ such that
$$
\int_{\ov B}|u_n|^2=1\ ,\quad \int_{Y\setminus \ov B}|u_n|^2+|\nabla u_n|^2\to0\ ,\quad \Delta u_n=0 \mbox{ in } \ov B.
$$
 Then $(u_n)$ is bounded in $W^{1,2}(Y)$ and its restriction to $Y\setminus \ov B$ does converges to $0$ strongly in $W^{1,2}(Y\setminus \ov B)$.
By exploiting (\ref{e.poincarre2}), we infer that this convergence to $0$ holds also in $W^{1,2}(Y)$ in contradiction with $\int_{\ov B}|u_n|^2=1$.

\vspace{0,5cm}
So far we have proved (\ref{e.poincarre2}) and (\ref{e.poincarre}) if $B(\theta,\rho)=B(\ov\theta, 1/2-\delta)$.
Let us consider the general case of $\theta\in Y$ and $\rho<1/2$ such that $\dist(\theta,\partial Y)\geq \rho+\delta$. It is easy to see that 
$$
B(\theta,\rho)=\theta+t\Big(\ov B-\big(\textstyle\frac12,\frac12\big)\Big) \quad\mbox{with }\quad  t=\frac{\rho}{\frac12-\delta}\ .
$$
With the same transformation, we define $Y_{\theta,\rho}:=\theta+t\Big(Y-\big(\frac12,\frac12\big)\Big)$. Clearly, by affine change of variable,
inequalities (\ref{e.poincarre2}) and (\ref{e.poincarre})  still hold true when we substitute $Y$ with $Y_{\theta,\rho}$ and $\ov B$ with $B(\theta, \rho)$ . For instance, the same constant $c_\delta$ appears provided the homothety factor $t$ satisfies $t\le 1$. 
The general case follows by observing that we have the inclusion\, $Y_{\theta,\rho}\setminus B(\theta,\rho)\subset \tor\setminus B(\theta,\rho)$.
\QED

\begin{lemma} \label{l.media}
Let  $c_\delta>0$ given in Lemma \ref{l.laplace} and $(\theta,\rho)$ with $\dist(\theta,\partial Y)\geq \rho+\delta$. Then for every 
$u\in W^{1,2}(Y)$ verifying
$\Delta u\in L^2( B(\theta,\rho))$, there holds
\begin{multline*}
\int_B |u|^2  \ \le\ \frac2{\mathrm{dist}^2\left(\alpha, \frac{\bsigma}{\rho^2}\right)} \int_{B(\theta,\rho)} |\Delta u + \alpha u|^2 \\ +
4\,c_\delta\, \left(1+\frac{|\alpha|^2}{\mathrm{dist}^2\big(\alpha,\frac{\bsigma}{\rho^2}\big)}\right)\, \ 
\int_{Y\setminus B(\theta,\rho)} (|u|^2 + |\nabla u|^2)\ .
\end{multline*}
\end{lemma}
\proof We set $B=B(\theta,\rho)$
and use for every  $u\in W^{1,2}(Y)$ the decomposition  $u=u_1+u_2$, where
\[
\Delta u_1=0\quad \text{on }B,\qquad u_1 = u \quad \text{on } Y\setminus B,\qquad u_2 := u-u_1\ .
\]
Then $u_2\in W^{1,2}_0(B)$ and as $\Delta u_2=\Delta u $ in the distribution sense on open subset $B$, 
we have that $u_2\in W^{2,2}(B)$ and by applying Lemma \ref{micro}
\begin{multline*}
 \mathrm{dist}\left(\alpha,\rho^{-2}\bsigma\right) \Vert u_2\Vert_{L^2(B)} \ \le\ \Vert\Delta u + \alpha (u- u_1)\Vert_{L^2(B)} 
\le  \Vert\Delta u + \alpha u\Vert_{L^2(B)} \, +\,  |\alpha|\, \Vert u_1\Vert_{L^2(B)}\ .
\end{multline*}
 Then by triangle inequality
 $$ \Vert u\Vert_{L^2(B)} \ \le\ \frac1{\mathrm{dist}\left(\alpha,\rho^{-2}\bsigma\right)} \Vert\Delta u + \alpha u\Vert_{L^2(B)} \, +\, 
 \left(1+ \frac{ |\alpha|}{\mathrm{dist}\left(\alpha,\rho^{-2}\bsigma\right)}\right)\, \Vert u_1\Vert_{L^2(B)}\ .$$
The conclusion follows by applying Lemma \ref{l.laplace} to $u_1$ noticing that $u_1=u$ on $Y\setminus B$.
\QED

\begin{remark}\upshape{
 In lemmas \ref{l.laplace} and \ref{l.media} the constant $c_\delta$ 
 blows up to infinity when $\delta\to0$.  We will apply both lemmas in 
 the infinitesimal domain $\eta Y$ in the form of the following rescaled inequalities 
\begin{equation}\label{e.poincarre_rescal} 
\int_{ \eta B(\theta,\rho)}\hskip-0.2cm |u|^2\leq c_\delta\int_{\eta (Y\setminus B(\theta,\rho))}\hskip-0.2cm |u|^2+\eta^2|\nabla u|^2 ,
\qquad
\int_{\eta B(\theta,\rho)}\hskip-0.2cm |\nabla u|^2\leq c_\delta\int_{\eta (Y\setminus B(\theta,\rho))} |\nabla u|^2 
\end{equation}
\begin{multline}
\int_{\eta B(\theta,\rho)}\hskip-0.2cm |u|^2  \ \le 4 \,c_\delta\, \left(1+\frac{|\alpha|^2}{\mathrm{dist}^2\big(\alpha,\frac{\bsigma}{\rho^2}\big)}\right)\, \ 
\int_{\eta (Y\setminus B(\theta,\rho))} (|u|^2 + \eta^2\, |\nabla u|^2)  \\
\label{poincare2_rescal}
+ \ \frac{2\,\eta^4}{\mathrm{dist}^2\left(\alpha, \frac{\bsigma}{\rho^2}\right)}\  \int_{\eta B(\theta,\rho)} |\Delta u + \frac{\alpha}{\eta^2} u|^2  
\end{multline}

}\end{remark}

\begin{lemma} \label{l.sup}
Let $\{X_i\, ,\, i=1,2,\dots, n,\dots\}$ be a sequence of identically distributed non negative random variables in $L^1(\Omega,\mathcal{A},\PPt)$.
Define \ $\disp Z_n:= \sup \{X_i, {1\leq i\leq n} \} .$
Then it holds:\  $\dis\frac{Z_n}{n} \ \Toas \ 0\ $
and  \ $\dis \lim_{n\to\infty}\frac1n \E(Z_n)=0\ . $  

\end{lemma}
\proof Let $\eps>0$. Noticing that the function $t\mapsto\PPt(\{X_n \ge t\eps\})$ is non increasing on $\Rset^+$ and does not depend on $n$, we have:
\begin{multline*}
\sum_1^\infty \PPt\left(\left\{\frac{X_n}{n}>\eps\right\}\right) =\sum_1^\infty \PPt\left(\left\{\frac{X_1}{n}>\eps\right\}\right)
\le \int_0^{+\infty} \PPt\left(\left\{X_1>t\eps\right\}\right) \, dt = \frac1{\eps} \, \E(X_1) \ <+\infty\ .
\end{multline*}
By Borel-Cantelli lemma, it follows that 
$\PPt\left( \limsup_n\left\{\frac{X_n}{n}>\eps\right\}\right)= 0$. Thus for a.a. $\omega$
there exists a finite integer $N(\omega)$ such that:
$$ \frac{Z_n(\omega)}{n} \ \le\ \max \left\{ \frac{Z_N(\omega)}{n}, \eps \right\}\ , \ \forall n\ge N(\omega).$$
We may assume that $Z_N(\omega)$ is finite. By letting $n\to\infty$ and then $\eps\to 0$, we conclude that $\dis\frac{Z_n}{n} \ \Toas \ 0.\ $
Furthermore, since $\{Z_n>t\} \subset \bigcup_{i=1}^n \{X_i> t\}$ hold for every $t$,
the function $f_n(t) :=\frac{\PPt(\{Z_n > t\})}{n}$ 
satisfies the upperbound $f_n(t) \le f(t):=\PPt(\{X_1>t\})$.
As $\int_0^\infty f(t)= \E(X_1)<+\infty$, we infer by dominated convergence that
$\dis \lim_{n\to\infty} \frac{\E(Z_n)}{n} \ =\ \lim_{n\to\infty} \int_0^\infty f_n(t)\, dt\ =\ 0\, .$ \QED

\bibliographystyle{plain}

\bibliography{random}
\end{document}